\documentclass[%
 reprint,
 amsmath,amssymb,
 aps,prl,
]{revtex4-1}

\usepackage{graphicx,placeins,subfig}
\usepackage{dcolumn}
\usepackage{bm}
\usepackage{xcolor}
\usepackage{physics}
\usepackage[normalem]{ulem}
\usepackage[final]{pdfpages}
\makeatletter
\AtBeginDocument{\let\LS@rot\@undefined}
\makeatother

\renewcommand{\ol}[1]{\overline{#1}}

\newcommand{\aop}{\hat{a}}
\newcommand{\cop}{{\aop^\dagger}}

\newcommand{\wv}[1]{#1}

\newcommand{\rr}[1]{#1}

\graphicspath{{submissionfigures/}}
\begin{document}

\preprint{APS/123-QED}

\title{Gaussian trajectory approach to dissipative phase transitions: the case of quadratically driven photonic lattices}
%

\author{Wouter Verstraelen}
\email{Wouter.Verstraelen@uantwerpen.be}
\affiliation{TQC, University of Antwerp, B-2610 Wilrijk, Belgium}

\author{Riccardo Rota}%
 \email{Riccardo.Rota@epfl.ch}
 \author{Vincenzo Savona}
\affiliation{Institute of Physics, Ecole Polytechnique F\'ed\'erale de Lausanne (EPFL), CH-1015 Lausanne, Switzerland}%
\author{Michiel Wouters}
\affiliation{TQC, University of Antwerp, B-2610 Wilrijk, Belgium}

\date{\today}

\begin{abstract}
We apply the Gaussian trajectories approach to the study of the critical behavior of two-dimensional dissipative arrays of nonlinear photonic cavities, in presence of two-photon driving and in regimes of sizable loss rates. In spite of the highly mixed character of the density matrix of this system, the numerical approach is able to provide precise estimations of the steady-state expectation values, even for large lattices made of more than 100 sites. By performing a finite-size scaling of the relevant properties of the steady state, we extrapolate the behavior of the system in the thermodynamic limit and we show the emergence of a second-order dissipative phase transition, belonging to the universality class of thermal Ising model. This result indicates the occurrence of a crossover when the loss rate is increased from the weak-loss limit, in which the phase transition belongs to the universality class of the quantum Ising model
\end{abstract}

\maketitle



Dissipative phase transitions are critical phenomena emerging in the non-equilibrium steady state of open quantum systems, due to the competition between their coherent Hamiltonian dynamics and the incoherent processes \cite{Kessler12,liouvillianSpectral}. In the last years, the possibility of realizing strongly correlated states in photonic cavity arrays \cite{qfloflight,Hartmann_2016,Noh_2016} has stimulated a deep investigation of these phenomena, which have been discussed theoretically in photonic systems \cite{Carmichael15,Mendoza-Arenas16,Casteels16,Bartolo16,Benito16,Casteels17_dimer,Casteels17_1stOrderDPT,Foss-Feig17,Biondi17,Biella17,VincenzoMF,Vicentini18,PRLspinmodel}, lossy polariton condensates \cite{Sieberer13,Sieberer14,Altman15} and spin models \cite{Lee11,Lee13,Chan15,Jin16,Maghrebi16,Rota17,Overbeck17,XYZdynamics,Roscher18}. From the experimental point of view, remarkable results have been recently obtained with driven circuit quantum electrodynamics systems \cite{Fink17} and semiconductor microcavities \cite{Rodriguez17,Fink2018}, showing the possibility to observe dissipative phase transitions in real open quantum many-body systems.

The non-trivial interplay between the Hamiltonian evolution, driving and dissipative processes in open quantum many-body systems has given rise to several fundamental questions about the respective roles  of quantum and classical fluctuations across a dissipative phase transition. In this regard, the universality classes of these critical phenomena has been intensely debated \cite{Diehl10,DallaTorrethermalno,DallaTorrethermalyes,Sieberer13,Sieberer14,Tauber14,Diehlnewuniversalityclass,Maghrebi16,Rota17}. In many cases, dissipative phase transitions emerge in regimes where the dissipation rates are comparable with the typical energy scale of the Hamiltonian and their universality classes belong to those of classical thermal phase transition \cite{Mitra06,DallaTorrethermalyes,DallaTorre13,Rota17}, such that the critical behavior can be described in term of an effective temperature emerging as a results of dissipations. 

In this debate, arrays of nonlinear photonic resonators in presence of two-photon -- i.e., quadratic in the field -— driving have recently attracted a certain attention, as these systems undergo a dissipative phase transition which belongs to a quantum universality class in a suitable regime of parameters \cite{PRLspinmodel}. Quadratically driven resonators have been realized experimentally with superconducting circuits \cite{Leghtas15,Wang16} and, in the last years,  several works have considered the possibility of exploiting the $\mathbb{Z}_2$ symmetry of this system in order to simulate the behavior of quantum spin systems, proposing also noise-resilient quantum codes based on this platform \cite{exacsolutionSciRep,Goto16,Goto2016,Nigg17,Puri2017}. In the context of critical phenomena, the spontaneous breaking of the $\mathbb{Z}_2$ symmetry in a lattice of coupled quadratically driven resonators has been investigated with both a mean-field \cite{VincenzoMF} and a many-body approach \cite{PRLspinmodel}, showing the emergence of a phase transition in a regime where the loss rate is small compared to the Hamiltonian energy scales, with critical exponents equal to those of the quantum transverse Ising model.

In this rapid, we consider the behavior of a quadratically driven photonic lattice when the loss rate becomes comparable with the Hamiltonian parameters, showing that the universality class of the transition changes in this regime and one recovers the description in terms of a classical criticality. We perform this study using the Gaussian trajectories approach, a novel numerical technique introduced in the study of the single Kerr cavity with a one-photon pump \cite{gaussianmethod}. This method turns out to be suitable to investigate dissipative phase transitions, as it allows to estimate efficiently the dissipative dynamics of large two-dimensional lattices ( of more than 100 sites) and, meanwhile, to include the relevant many-body correlations which arise in the vicinity of the critical point. Analyzing the finite-size scaling of the steady-state properties of the system, we estimate its behavior in the thermodynamic limit of infinitely large lattices and highlight the emergence of a second-order phase transition, with critical exponents equal to those of the classical Ising model.

A lattice of $N$ coupled photonic cavities can be described by a Bose-Hubbard model, whose Hamiltonian contains a local term $\hat{h}_i$ acting on the $i$-th cavity, and a non-local term modelling the photon hopping between different cavities (we set $\hbar = 1$):
\begin{equation}\label{eq:Hamiltonian}
    \hat{H}=\sum_{i=1}^N \hat{h}_i-\sum_{\langle i j\rangle}\frac{J}{z}(\cop_i\aop_j+\cop_j\aop_i) \ .
\end{equation}
Here, $\aop_i (\cop_i)$ is the annihilation (creation) operator acting on the $i$-th site and the last sum runs over the nearest-neighbor pairs $\langle i j \rangle$, $J$ and $z$ being, respectively, the hopping strength and the coordination number. The local term can be written as 
\begin{equation}
    \hat{h}_i=-\Delta\cop_i\aop_i+\frac{U}{2}\aop_i^{\dagger 2}\aop_i^2+\frac{G}{2}{\aop_i}^{\dagger 2}+\frac{G^*}{2}\aop_i^2 \ ,
\end{equation}
where $\Delta$ is the detuning between half of the two-photon driving field frequency and the resonant cavity frequency, $U$ is the photon-photon interaction energy associated to the Kerr nonlinearity and $G$ is the two-photon driving field amplitude.

Assuming that the dissipative processes are Markovian, the dynamics of the open quantum system is recovered in terms of the density matrix $\hat{\rho}(t)$ which solves the Lindblad master Equation:
\begin{equation}\label{eq:Liouvillian}
    \pdv{\hat{\rho}}{t}=\mathcal{L}\hat{\rho},=-i\comm{\hat{H}}{\hat{\rho}}+\sum_{j,k}\hat{\Gamma}_{j,k}\hat{\rho}\hat{\Gamma}_{j,k}^\dagger-\frac{1}{2}\acomm{\hat{\Gamma}_{j,k}^\dagger\hat{\Gamma}_{j,k}}{\hat{\rho}} \ .
\end{equation}
Here, $\mathcal{L}$ is the Liouvillian superoperator and $\hat{\Gamma}_{j,k}$ are the jump operators which describe the coupling of the system with the environment: in general, the system exhibits local one- and two-photon losses, which are modelled respectively with the jump operators $\hat{\Gamma}_{1,j}=\sqrt{\gamma}\aop_j$ and $\hat{\Gamma}_{2,j}=\sqrt{\eta}\aop_j^2$. 
\wv{In particular, we will assume that the operators $\hat{\Gamma}_{1,j}$ ensure that the system relaxes towards a nonequilibrium steady state $\hat{\rho}_{SS}$, and the emergence of a persistent non-stationarity \cite{Buca2019,Albert2016} is avoided.  Conditions for the existence of a unique steady state have been largely investigated in the literature \cite{Spohn1976,*Spohn1977,Nigro_2019}.}

Liouvillian $\mathcal{L}$ presents a $\mathbb{Z}_2$ symmetry coming from its invariance under a global change of sign of the annihilation operators, $\aop_i \to -\aop_i \ \forall i$. In the thermodynamic limit of an infinite lattice, the $\mathbb{Z}_2$ symmetry can be spontaneously broken for large values of $G/\gamma$ and $J/\gamma$, as indicated by a mean-field analysis \cite{VincenzoMF}. This leads to the emergence of a transition between a phase with a $\mathbb{Z}_2$-symmetric steady state (i.e. $\textrm{Tr}(\hat{\rho}_{SS} \sum_i \aop_i) = 0$) and a coherent phase with non-zero expectation value of the Bose field (i.e. $\textrm{Tr}(\hat{\rho}_{SS} \sum_i \aop_i) \ne 0$).

The nature of the dissipative phase transition can be investigated in terms of an approximate spin model, where two Schr\"{o}dinger-cat states with opposite parity play the role of the two $s=1/2$-spin states with opposite magnetization \cite{PRLspinmodel}. When projecting the bosonic Hamiltonian in Eq. \eqref{eq:Hamiltonian} onto the basis spanned by the Schr\"{o}dinger-cat states, we recover the Hamiltonian of a quantum transverse XY model, where the photon hopping term plays the role of an anisotropic spin coupling in the $xy$ plane (the coupling can be either ferromagnetic \cite{PRLspinmodel} or antiferromagnetic \cite{antiferromagnet}, according to the sign of $J$) and the detuning $\Delta$ plays the role of a transverse magnetic field in the $z$-direction. This model is known to present a phase transition belonging to the universality class of the quantum transverse Ising model \cite{dutta2015}.

From a computational perspective, the numerical solution of master equation \eqref{eq:Liouvillian} becomes intractable even for small lattices. The corner-space renormalization method \cite{cornerspace}, has been used to study quadratically driven-dissipative lattices with large nonlinearities and small loss rates, but it fails in the description of regimes characterized by a large photon occupancy per cavity and a highly mixed steady state, which are the regimes we consider here.

An alternative method which can overcome these difficulties is the Gaussian Trajectories approach (GTA), which has been developed for the numerical simulation of polaritonic systems \cite{gaussianmethod} and applied to study the temporal coherence of a dye-microcavity photon condensate \cite{photoncondensate}. As in an exact Wave Function Monte Carlo approach \cite{MolmerJOSAB93}, the GTA recovers the density matrix $\hat{\rho}(t)$ of the open quantum system from the average of a set of $N_T$ pure states $\ket{\psi_n(t)}$ (usually called \textit{quantum trajectories}) obtained independently by integrating a stochastic differential equation: $\hat{\rho}(t) = \frac{1}{N_T} \sum_{j=1}^{N_T} \ket{\psi_n(t)}\bra{\psi_n(t)}$. The physical picture underlying this formalism is to consider the environment as a measurement apparatus continuously monitoring the open system, and hence the stochastic evolution of each quantum trajectory may be interpreted as the result of the different random outcomes of these measurements \cite{haroche_raimond_2013,CarmichaelBOOK,WisemanBOOK,Daley14}. The peculiarity of the GTA approach is the Gaussian ansatz for the quantum trajectories, which makes the pure state $\ket{\psi_n}$ completely determined by its first and second central moments \cite{GaussquantinfRMP,GaussQInf}. This assumption reduces notably the computational cost needed for the numerical integration of the stochastic differential equation, since the complexity of the problem within GTA scales quadratically with the number $N$ of lattice sites, rather than exponentially.

Contrarily to the exact Wave Function Monte Carlo method, where there is no univocal choice of the stochastic differential equation for the trajectories $\ket{\psi_n(t)}$ in order to reproduce the same master equation for the density matrix $\hat{\rho}(t)$, the assumption of a Gaussian ansatz for $\ket{\psi_n(t)}$ makes the accuracy of the method dependent on the choice of the stochastic \emph{unravelings} used to factorize the coupling of the quantum system with the external environment.
For the single site problem with quadratic driving and losses, it was shown in \cite{2photdrivingunraveling}, that in the homodyne unraveling \cite{quantumnoise}, exact trajectories remain nearly coherent states, a subclass of Gaussian states. Here, we will therefore use the heterodyne unraveling, a symmetrized version of homodyne detection.
Heterodyne measurements measure the field quadratures with equal weights and are equivalent to two complementary homodyne detection schemes. It is a common procedure in quantum optics, where the  photon losses are interfered with an out-of-resonance reference beam \cite{quantumnoise} and is mathematically closely connected to the stochastic collapse model known as \emph{Quantum State Diffusion}(QSD)\cite{breuer}. 
 Interestingly, it has recently been suggested that such unraveling naturally captures macroscopic phenomena as phase transitions \cite{Strunz_1998,phasetransitionwithtrajectories}.
 The derivation of the full stochastic differential equations for the dynamics of the quadratically driven-dissipative photonic lattice within the GTA can be found in Supplemental material \cite{sup}.

\rr{Another aspect which makes the GTA suitable for the study of dissipative phase transitions} is that the \rr{weak} $\mathbb{Z}_2$-symmetry of the Liouvillian is not preserved in a single quantum trajectory, but it is restored only after averaging many trajectories. Therefore, it is possible to obtain non-trivial results in the calculation of a suitable order parameter along an individual trajectory at long times. The study of the distribution of the latter expectation values over the whole set of sampled trajectories is helpful to understand the emergence of collective phases in different regimes of the physical parameters, and hence provides an important insight into the critical behavior in the thermodynamic limit \cite{XYZdynamics}.

We apply the GTA formalism to the study of the quadratically-driven dissipative Bose-Hubbard model in 2D square lattices of different size, with periodic boundary conditions. In Supplemental material \cite{sup}, we show that the method is able to replicate the finding of quantum-critical behavior for parameters similar to \cite{PRLspinmodel}, \rr{and furthermore a benchmark study that shows the accuracy of this method for a dimer}. Since our main objective is to investigate the dissipative phase transition in regimes where the loss rates are comparable with the Hamiltonian parameters, we here set $U/\gamma=J/\gamma=1$, $\Delta = -J$. The last condition assures that the two-photon driving is resonant with the $k=0$ mode of the single-particle energy band of the closed system, and is useful to avoid any bistable behavior. Moreover, since two-photon losses are not expected to play an important role in the emergent criticality \cite{PRLspinmodel}, we set $\eta = 0$: this choice allows for a relevant speed-up of the numerical calculations (\cite{sup}).

\begin{figure}
    \centering
    \includegraphics[width=0.85\linewidth]{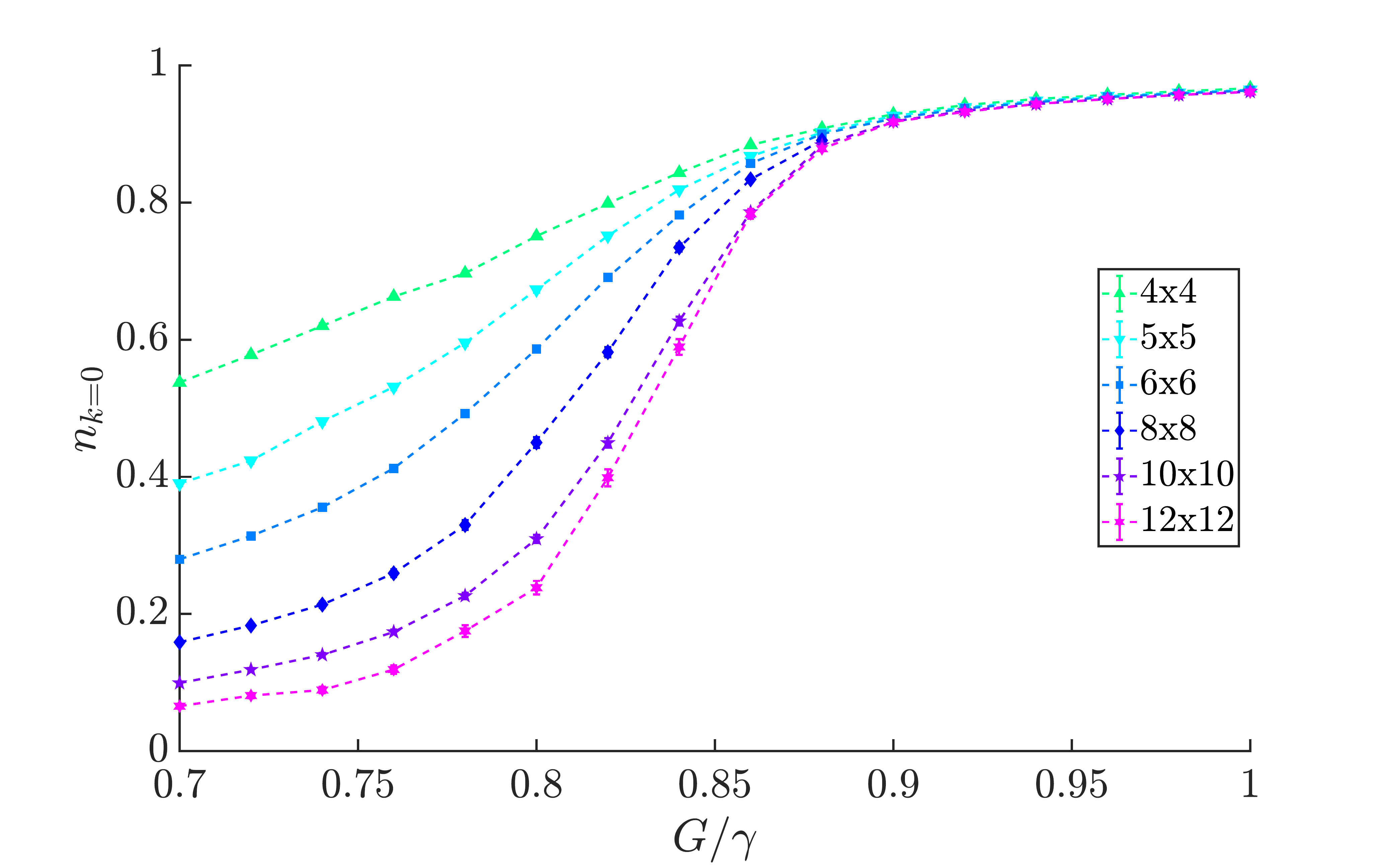}
    \caption{The steady-state relative occupation of the $k=0$-mode increases as a function $G/\gamma$, as computed in lattices of different size. Results are obtained from 1000 trajectories (500 for the 12x12 lattice), each evolved over a total time $t = 100\gamma^{-1}$. For each quantum trajectory, the vacuum is chosen as initial state and the two-photon driving is switched on slowly to the desired value, in order to reduce the formation of defects which would appear from strong quenching \cite{KZreview}. The steady-state expectation values are computed averaging the results obtained in a time interval from a size-dependent initial time $t_\text{relax}$ to the final time $t_{fin} = 100\gamma^{-1}$.
    }
    \label{fig:k0occup}
\end{figure}

In Fig. \ref{fig:k0occup}, the steady-state expectation value for the relative occupation of the $k=0$ mode
\begin{equation}
n_{k=0}=\frac{\sum_{jj'}\textrm{Tr}\left(\hat{\rho}_{SS}\cop_j\aop_{j'}\right)}{\left[\sum_{j}\textrm{Tr}\left(\hat{\rho}_{SS}\cop_j\aop_{j}\right)\right]^2}
\end{equation}
is plotted as a function of the driving amplitude $G$ for different lattice sizes. At large values of $G/\gamma$, we notice that $n_{k=0} \simeq 1$, independently of the size of the lattice. This result indicates that the steady state of the dissipative system in this limit is a coherent state characterized by long-range order, since the local Bose fields on each cavity assume the same value. This is in agreement with the picture of the spontaneous symmetry breaking obtained the mean-field calculation \cite{VincenzoMF}. At smaller values of $G/\gamma$, $n_{k=0}$ decreases when the lattice size increases, suggesting that $n_{k=0} \to 0$ in the thermodynamic limit. This behavior supports the hypothesis of the presence of a disordered $\mathbb{Z}_2$-symmetric phase for small values of the driving amplitude.

\begin{figure}
    \centering
\includegraphics[width=0.9\linewidth]{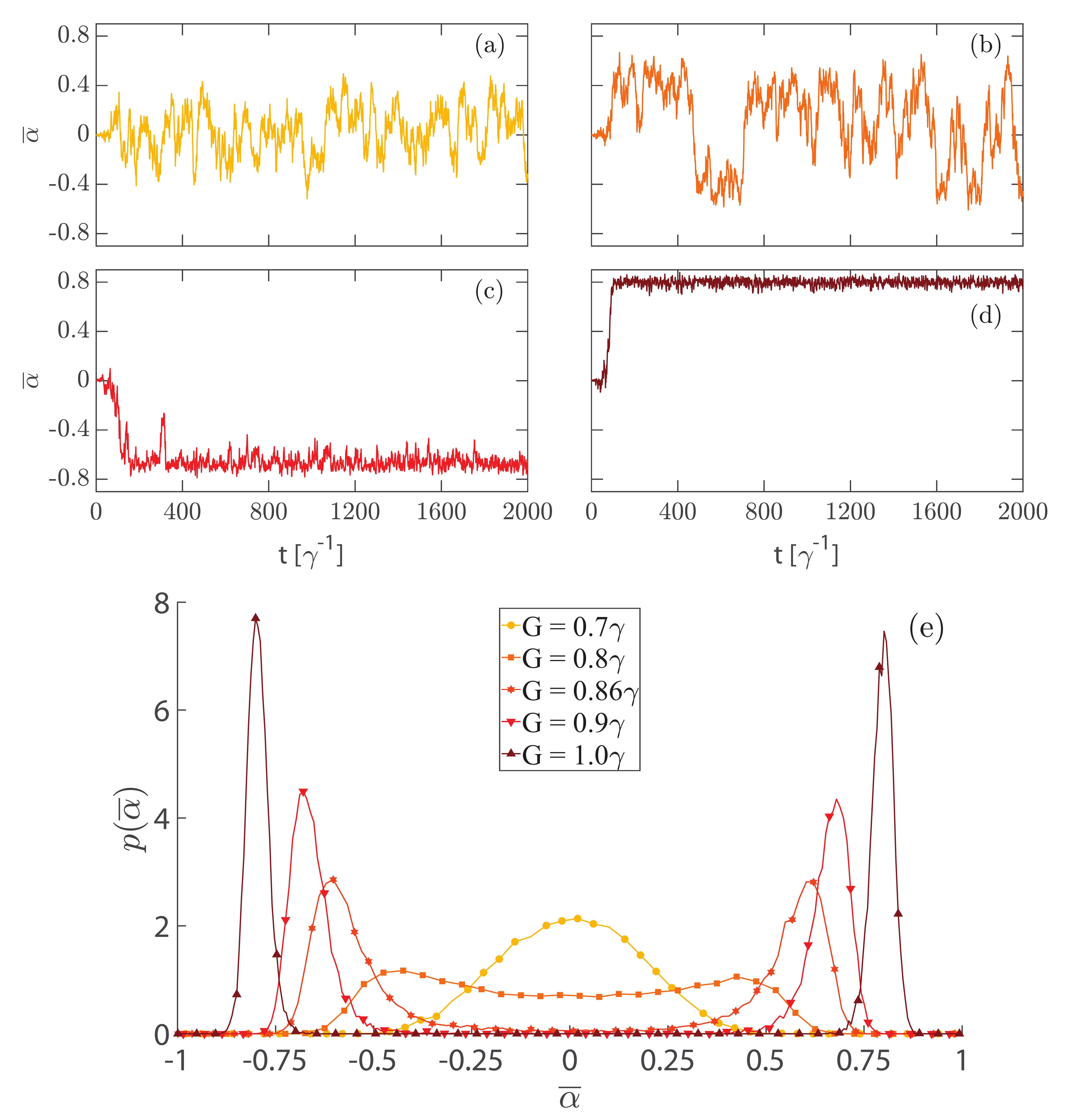}    
    \caption{(a-d): Time evolution of the order parameter $\ol{\alpha}$ along a single Gaussian trajectory in a $6\times6$ lattice, for different values of the driving amplitude: (a) $G/\gamma = 0.7$, (b) $G/\gamma = 0.8$, (c) $G/\gamma = 0.9$, (d) $G/\gamma = 1.0$. (e) Probability distribution $p(\ol{\alpha})$ of the order parameter for different values of $G$. For each value of $G$, we consider 1000 trajectories each evolved over a time $t = 400$ and we collect the data for $\ol{\alpha}(t)$ for $t\ge 20$, i.e. for long times in which the density matrix of the dissipative system has reached the steady state. The distribution $p(\ol{\alpha})$ is obtained as the histogram of the collected data. As $G$ increases, a qualitative transition from monomodal to bimodal distribution is evident.} 

    \label{fig:singleshots}
\end{figure}

Given this last result, it is convenient to describe the behavior of the system across the phase transition in terms of an order parameter, which is related to the average value of the Bose fields on the different cavities. We thus define $\ol{\alpha} = \Im \bra{\psi}\frac{1}{N}\sum_i\aop_i\ket{\psi}$, i.e. the expectation value of the average Bose field on a single quantum trajectory (we consider its imaginary part in order to have a real-valued order parameter): this quantity has a clear physical meaning, as it corresponds to the outcome of a heterodyne unraveling in real experiments. 
In Fig. \ref{fig:singleshots}-(a-d), we plot the time evolution of $\overline{\alpha}$ on single trajectories in a $6\times6$ lattice, choosing the vacuum as the initial state at $t=0$, for different values of the driving amplitude. For $G=0.7 \gamma$ [Fig. \ref{fig:singleshots}-(a)], we notice a behavior supporting the emergence of a disordered phase, as the order parameter exhibits Gaussian fluctuations around the value $\ol{\alpha}=0$. For $G=0.8\gamma$ [Fig. \ref{fig:singleshots}-(b)], the situation is similar, but we can see the presence of longer intervals of time where the value of $\ol{\alpha}$ stagnates around a non-zero values, suggesting the appearance of a bimodal character in the steady-state distribution of the order parameter. At larger values of the driving amplitude [$G \ge 0.9\gamma$, Fig. \ref{fig:singleshots}-(c-d)], the behavior of $\ol{\alpha}(t)$ is completely different: after a first transient where the value of $\ol{\alpha}$ changes notably in time, at long time we see that the order parameter fluctuates around a non-zero value, which can be either negative [as in Fig. \ref{fig:singleshots}-(c)] or positive [as in Fig. \ref{fig:singleshots}-(d)] according to the particular realization of the noise in the stochastic differential equation. This behavior suggest the emergence of an ordered phase with broken symmetry in this regime of parameters. However, when considering an ensemble of many trajectories at given $G$, half of them will stabilize at long time around a positive value for $\ol{\alpha}$ and the other half around the opposite value $-\ol{\alpha}$, retrieving thus a $\mathbb{Z}_2$-symmetric steady-state density matrix, as expected for a system of finite size. This behavior becomes clear from sampling $p(\ol{\alpha})$, the probability distribution of $\ol{\alpha}$, is shown in Fig. \ref{fig:singleshots}-(e). Although $p(\ol{\alpha})$ is always an even function, we notice that, when increasing $G$, the character of the distribution changes from a monomodal to a bimodal behavior.

\begin{figure}
    \centering
    \includegraphics[width=0.9\linewidth]{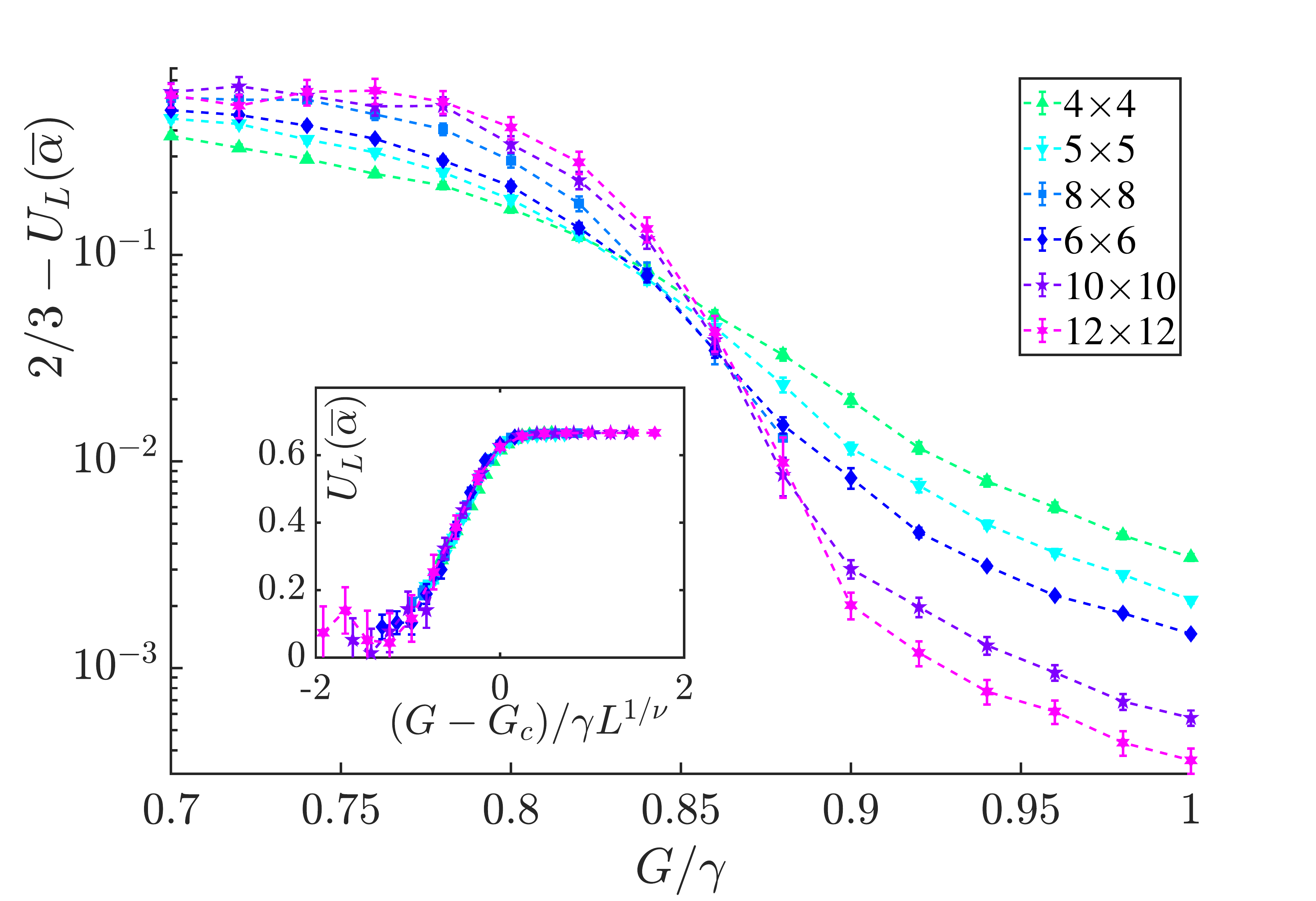}
    \caption{Binder cumulant $U_L$ as a function of the two-photon driving amplitude $G$, for different lattice sizes $L$. A crossing of the curves for different lattice sizes can be seen, signaling the emergence of a critical point at $G = G_c$ (the difference $2/3-U_L$ is shown on a logarithmic scale). Inset: The curves for different sizes show an universal behavior when plotted as a function of $(G - G_c) L^{1/\nu}$, with $\nu=1$ the critical exponent for the correlation length in the classical 2D Ising model}
    \label{fig:binder}
\end{figure}

A more quantitative analysis can be performed from the Binder cumulant, defined as $U_L=1-\mu_4/(3\mu_2^2)$,
where $\mu_m = \int d\ol{\alpha} \, \ol{\alpha}^m \, p(\ol{\alpha})$ denotes the $m$-th moment of the probability distribution $p(\ol{\alpha})$ \cite{Binder1981,binderpar}. This quantity has been deeply used in the finite-size scaling analysis of spin models in presence of a paramagnetic-to-ferromagnetic phase transition, thanks to its peculiarity of being a universal function of the ratio $\xi/L$, with $\xi$ and $L$ being respectively the correlation length and the finite size of the lattice \cite{Binder1981}. This means that at a critical point, where the correlation length diverges in the thermodynamic limit, the Binder cumulant becomes independent of the lattice size $L$: therefore, the finite-size scaling analysis of $U_L$ is a useful approach to track the emergence of the criticality in  our system. In Fig. \ref{fig:binder}, we plot $U_L$ as a function of the driving amplitude $G$ for lattices of different size. We find that all the different curves cross at the common point $G_c = 0.86 \gamma$, confirming the presence of a dissipative phase transition in our model. Some insight about the universality class of the phase transition can be obtained by plotting the same data for $U_L(\overline{\alpha})$ as a function of $(G-G_c) L^{1/\nu}$, where $\nu$ is the critical exponent of the correlation length. From the results shown in Fig. \ref{fig:binder} (inset), we notice that setting $\nu = 1$ makes all the curves collapse, confirming the expected universal behavior of $U_L = f(\xi/L)$. The value $\nu = 1$ corresponds to the critical exponent of the correlation length for a 2D \emph{classical} Ising model. The latter result indicates that the dissipative phase transition of a quadratically-driven Bose-Hubbard model in the regime of large loss rates results from a classical criticality, contrarily to what happens in regimes of small loss rate where the phase transition has a quantum nature \cite{PRLspinmodel}.

The GTA method provides a simple picture of the crossover between quantum and classical critical behavior: whereas quantum correlations between sites $n$ and $m$ are described by co\"efficients $u_{nm}$ and $v_{nm}$, classical correlations are given by the statistical correlator $\ev{\alpha_n \alpha_m}_s$, where$\ev{\cdot}_s$ denotes an average over trajectories. It is clear that strong dissipation has the same effect as strong measurements: it transfers quantum correlations to classical ones\cite{breuer}.

In this work, we have applied the Gaussian trajectory approach in the study of quadratically driven photonic lattices across a dissipative phase transition. This method turns out to be particularly suitable for theoretical analysis of critical phenomena in open quantum systems, as it provides accurate estimates of the physical observables even in regimes where the steady state is highly mixed.

In the case under consideration, we have been able to simulate lattices of up to 144 sites, which have allowed to perform a precise finite-size scaling of the relevant properties of the system and to extract the critical exponent $\nu$ for the correlation length. The result $\nu = 1$ is found, which differs from the value predicted by mean-field theory and corresponds to the classical Ising model, revealing the ability of the method to describe the many-body correlations arising among the photons. Furthermore, in the low-loss regime (\cite{sup}) we replicate the finding of the value $\nu=\nu_q$  of the quantum Ising model which demonstrates that the method is able to capture the entanglement leading to relevant quantum correlations. This suggests a scenario in which, similarly to the case at thermal equilibrium, a crossover between quantum and classical criticality in the non-equilibrium steady state occurs, depending on the scale of the loss rate relative to the Hamiltonian energy scale.

The possibility to use the Gaussian ansatz in the description of strongly correlated photons can open up several ways to the application of our method in quantum many-body physics with light. An interesting perspective could be to use Gaussian trajectories to investigate the emergence of collective phenomena in photonic lattices in presence of geometric frustration \cite{antiferromagnet,Biondi18} or disorder \cite{Vicentini19}. Furthermore, the method is readily applicable to the study of the dynamics is these systems: \rr{very recently, we have studied dynamical aspects of the same phase transition with the GTA method in Ref. \cite{dynamicalPT}}.

Our case study to the quadratically driven kerr lattice may also contribute to the general understanding of the crossover from quantum to classical critical behavior in dissipative phase transitions.

\FloatBarrier
\begin{acknowledgments}
We acknowledge stimulating discussions with F. Minganti and D. Huybrechts. W.V. acknowledges financial support from the FWO project 41190, the FWO travel grant V413119N and the hospitality of LTPN.
\end{acknowledgments}



\begin{thebibliography}{76}%
\makeatletter
\providecommand \@ifxundefined [1]{%
 \@ifx{#1\undefined}
}%
\providecommand \@ifnum [1]{%
 \ifnum #1\expandafter \@firstoftwo
 \else \expandafter \@secondoftwo
 \fi
}%
\providecommand \@ifx [1]{%
 \ifx #1\expandafter \@firstoftwo
 \else \expandafter \@secondoftwo
 \fi
}%
\providecommand \natexlab [1]{#1}%
\providecommand \enquote  [1]{``#1''}%
\providecommand \bibnamefont  [1]{#1}%
\providecommand \bibfnamefont [1]{#1}%
\providecommand \citenamefont [1]{#1}%
\providecommand \href@noop [0]{\@secondoftwo}%
\providecommand \href [0]{\begingroup \@sanitize@url \@href}%
\providecommand \@href[1]{\@@startlink{#1}\@@href}%
\providecommand \@@href[1]{\endgroup#1\@@endlink}%
\providecommand \@sanitize@url [0]{\catcode `\\12\catcode `\$12\catcode
  `\&12\catcode `\#12\catcode `\^12\catcode `\_12\catcode `\%12\relax}%
\providecommand \@@startlink[1]{}%
\providecommand \@@endlink[0]{}%
\providecommand \url  [0]{\begingroup\@sanitize@url \@url }%
\providecommand \@url [1]{\endgroup\@href {#1}{\urlprefix }}%
\providecommand \urlprefix  [0]{URL }%
\providecommand \Eprint [0]{\href }%
\providecommand \doibase [0]{https://doi.org/}%
\providecommand \selectlanguage [0]{\@gobble}%
\providecommand \bibinfo  [0]{\@secondoftwo}%
\providecommand \bibfield  [0]{\@secondoftwo}%
\providecommand \translation [1]{[#1]}%
\providecommand \BibitemOpen [0]{}%
\providecommand \bibitemStop [0]{}%
\providecommand \bibitemNoStop [0]{.\EOS\space}%
\providecommand \EOS [0]{\spacefactor3000\relax}%
\providecommand \BibitemShut  [1]{\csname bibitem#1\endcsname}%
\let\auto@bib@innerbib\@empty
\bibitem [{\citenamefont {Kessler}\ \emph {et~al.}(2012)\citenamefont
  {Kessler}, \citenamefont {Giedke}, \citenamefont {Imamoglu}, \citenamefont
  {Yelin}, \citenamefont {Lukin},\ and\ \citenamefont {Cirac}}]{Kessler12}%
  \BibitemOpen
  \bibfield  {author} {\bibinfo {author} {\bibfnamefont {E.~M.}\ \bibnamefont
  {Kessler}}, \bibinfo {author} {\bibfnamefont {G.}~\bibnamefont {Giedke}},
  \bibinfo {author} {\bibfnamefont {A.}~\bibnamefont {Imamoglu}}, \bibinfo
  {author} {\bibfnamefont {S.~F.}\ \bibnamefont {Yelin}}, \bibinfo {author}
  {\bibfnamefont {M.~D.}\ \bibnamefont {Lukin}},\ and\ \bibinfo {author}
  {\bibfnamefont {J.~I.}\ \bibnamefont {Cirac}},\ }\bibfield  {title} {\bibinfo
  {title} {Dissipative phase transition in a central spin system},\ }\href
  {https://doi.org/10.1103/PhysRevA.86.012116} {\bibfield  {journal} {\bibinfo
  {journal} {Phys. Rev. A}\ }\textbf {\bibinfo {volume} {86}},\ \bibinfo
  {pages} {012116} (\bibinfo {year} {2012})}\BibitemShut {NoStop}%
\bibitem [{\citenamefont {Minganti}\ \emph {et~al.}(2018)\citenamefont
  {Minganti}, \citenamefont {Biella}, \citenamefont {Bartolo},\ and\
  \citenamefont {Ciuti}}]{liouvillianSpectral}%
  \BibitemOpen
  \bibfield  {author} {\bibinfo {author} {\bibfnamefont {F.}~\bibnamefont
  {Minganti}}, \bibinfo {author} {\bibfnamefont {A.}~\bibnamefont {Biella}},
  \bibinfo {author} {\bibfnamefont {N.}~\bibnamefont {Bartolo}},\ and\ \bibinfo
  {author} {\bibfnamefont {C.}~\bibnamefont {Ciuti}},\ }\bibfield  {title}
  {\bibinfo {title} {Spectral theory of liouvillians for dissipative phase
  transitions},\ }\href {https://doi.org/10.1103/PhysRevA.98.042118} {\bibfield
   {journal} {\bibinfo  {journal} {Phys. Rev. A}\ }\textbf {\bibinfo {volume}
  {98}},\ \bibinfo {pages} {042118} (\bibinfo {year} {2018})}\BibitemShut
  {NoStop}%
\bibitem [{\citenamefont {Carusotto}\ and\ \citenamefont
  {Ciuti}(2013)}]{qfloflight}%
  \BibitemOpen
  \bibfield  {author} {\bibinfo {author} {\bibfnamefont {I.}~\bibnamefont
  {Carusotto}}\ and\ \bibinfo {author} {\bibfnamefont {C.}~\bibnamefont
  {Ciuti}},\ }\bibfield  {title} {\bibinfo {title} {Quantum fluids of light},\
  }\href {https://doi.org/10.1103/RevModPhys.85.299} {\bibfield  {journal}
  {\bibinfo  {journal} {Rev. Mod. Phys.}\ }\textbf {\bibinfo {volume} {85}},\
  \bibinfo {pages} {299} (\bibinfo {year} {2013})}\BibitemShut {NoStop}%
\bibitem [{\citenamefont {Hartmann}(2016)}]{Hartmann_2016}%
  \BibitemOpen
  \bibfield  {author} {\bibinfo {author} {\bibfnamefont {M.~J.}\ \bibnamefont
  {Hartmann}},\ }\bibfield  {title} {\bibinfo {title} {Quantum simulation with
  interacting photons},\ }\href
  {https://doi.org/10.1088/2040-8978/18/10/104005} {\bibfield  {journal}
  {\bibinfo  {journal} {Journal of Optics}\ }\textbf {\bibinfo {volume} {18}},\
  \bibinfo {pages} {104005} (\bibinfo {year} {2016})}\BibitemShut {NoStop}%
\bibitem [{\citenamefont {Noh}\ and\ \citenamefont
  {Angelakis}(2016)}]{Noh_2016}%
  \BibitemOpen
  \bibfield  {author} {\bibinfo {author} {\bibfnamefont {C.}~\bibnamefont
  {Noh}}\ and\ \bibinfo {author} {\bibfnamefont {D.~G.}\ \bibnamefont
  {Angelakis}},\ }\bibfield  {title} {\bibinfo {title} {Quantum simulations and
  many-body physics with light},\ }\href
  {https://doi.org/10.1088/0034-4885/80/1/016401} {\bibfield  {journal}
  {\bibinfo  {journal} {Reports on Progress in Physics}\ }\textbf {\bibinfo
  {volume} {80}},\ \bibinfo {pages} {016401} (\bibinfo {year}
  {2016})}\BibitemShut {NoStop}%
\bibitem [{\citenamefont {Carmichael}(2015)}]{Carmichael15}%
  \BibitemOpen
  \bibfield  {author} {\bibinfo {author} {\bibfnamefont {H.~J.}\ \bibnamefont
  {Carmichael}},\ }\bibfield  {title} {\bibinfo {title} {Breakdown of photon
  blockade: A dissipative quantum phase transition in zero dimensions},\ }\href
  {https://doi.org/10.1103/PhysRevX.5.031028} {\bibfield  {journal} {\bibinfo
  {journal} {Phys. Rev. X}\ }\textbf {\bibinfo {volume} {5}},\ \bibinfo {pages}
  {031028} (\bibinfo {year} {2015})}\BibitemShut {NoStop}%
\bibitem [{\citenamefont {Mendoza-Arenas}\ \emph {et~al.}(2016)\citenamefont
  {Mendoza-Arenas}, \citenamefont {Clark}, \citenamefont {Felicetti},
  \citenamefont {Romero}, \citenamefont {Solano}, \citenamefont {Angelakis},\
  and\ \citenamefont {Jaksch}}]{Mendoza-Arenas16}%
  \BibitemOpen
  \bibfield  {author} {\bibinfo {author} {\bibfnamefont {J.~J.}\ \bibnamefont
  {Mendoza-Arenas}}, \bibinfo {author} {\bibfnamefont {S.~R.}\ \bibnamefont
  {Clark}}, \bibinfo {author} {\bibfnamefont {S.}~\bibnamefont {Felicetti}},
  \bibinfo {author} {\bibfnamefont {G.}~\bibnamefont {Romero}}, \bibinfo
  {author} {\bibfnamefont {E.}~\bibnamefont {Solano}}, \bibinfo {author}
  {\bibfnamefont {D.~G.}\ \bibnamefont {Angelakis}},\ and\ \bibinfo {author}
  {\bibfnamefont {D.}~\bibnamefont {Jaksch}},\ }\bibfield  {title} {\bibinfo
  {title} {Beyond mean-field bistability in driven-dissipative lattices:
  Bunching-antibunching transition and quantum simulation},\ }\href
  {https://doi.org/10.1103/PhysRevA.93.023821} {\bibfield  {journal} {\bibinfo
  {journal} {Phys. Rev. A}\ }\textbf {\bibinfo {volume} {93}},\ \bibinfo
  {pages} {023821} (\bibinfo {year} {2016})}\BibitemShut {NoStop}%
\bibitem [{\citenamefont {Casteels}\ \emph {et~al.}(2016)\citenamefont
  {Casteels}, \citenamefont {Storme}, \citenamefont {Le~Boit\'e},\ and\
  \citenamefont {Ciuti}}]{Casteels16}%
  \BibitemOpen
  \bibfield  {author} {\bibinfo {author} {\bibfnamefont {W.}~\bibnamefont
  {Casteels}}, \bibinfo {author} {\bibfnamefont {F.}~\bibnamefont {Storme}},
  \bibinfo {author} {\bibfnamefont {A.}~\bibnamefont {Le~Boit\'e}},\ and\
  \bibinfo {author} {\bibfnamefont {C.}~\bibnamefont {Ciuti}},\ }\bibfield
  {title} {\bibinfo {title} {Power laws in the dynamic hysteresis of quantum
  nonlinear photonic resonators},\ }\href
  {https://doi.org/10.1103/PhysRevA.93.033824} {\bibfield  {journal} {\bibinfo
  {journal} {Phys. Rev. A}\ }\textbf {\bibinfo {volume} {93}},\ \bibinfo
  {pages} {033824} (\bibinfo {year} {2016})}\BibitemShut {NoStop}%
\bibitem [{\citenamefont {Bartolo}\ \emph {et~al.}(2016)\citenamefont
  {Bartolo}, \citenamefont {Minganti}, \citenamefont {Casteels},\ and\
  \citenamefont {Ciuti}}]{Bartolo16}%
  \BibitemOpen
  \bibfield  {author} {\bibinfo {author} {\bibfnamefont {N.}~\bibnamefont
  {Bartolo}}, \bibinfo {author} {\bibfnamefont {F.}~\bibnamefont {Minganti}},
  \bibinfo {author} {\bibfnamefont {W.}~\bibnamefont {Casteels}},\ and\
  \bibinfo {author} {\bibfnamefont {C.}~\bibnamefont {Ciuti}},\ }\bibfield
  {title} {\bibinfo {title} {Exact steady state of a kerr resonator with one-
  and two-photon driving and dissipation: Controllable wigner-function
  multimodality and dissipative phase transitions},\ }\href
  {https://doi.org/10.1103/PhysRevA.94.033841} {\bibfield  {journal} {\bibinfo
  {journal} {Phys. Rev. A}\ }\textbf {\bibinfo {volume} {94}},\ \bibinfo
  {pages} {033841} (\bibinfo {year} {2016})}\BibitemShut {NoStop}%
\bibitem [{\citenamefont {Benito}\ \emph {et~al.}(2016)\citenamefont {Benito},
  \citenamefont {S\'anchez Mu\~noz},\ and\ \citenamefont
  {Navarrete-Benlloch}}]{Benito16}%
  \BibitemOpen
  \bibfield  {author} {\bibinfo {author} {\bibfnamefont {M.}~\bibnamefont
  {Benito}}, \bibinfo {author} {\bibfnamefont {C.}~\bibnamefont {S\'anchez
  Mu\~noz}},\ and\ \bibinfo {author} {\bibfnamefont {C.}~\bibnamefont
  {Navarrete-Benlloch}},\ }\bibfield  {title} {\bibinfo {title} {Degenerate
  parametric oscillation in quantum membrane optomechanics},\ }\href
  {https://doi.org/10.1103/PhysRevA.93.023846} {\bibfield  {journal} {\bibinfo
  {journal} {Phys. Rev. A}\ }\textbf {\bibinfo {volume} {93}},\ \bibinfo
  {pages} {023846} (\bibinfo {year} {2016})}\BibitemShut {NoStop}%
\bibitem [{\citenamefont {Casteels}\ and\ \citenamefont
  {Ciuti}(2017)}]{Casteels17_dimer}%
  \BibitemOpen
  \bibfield  {author} {\bibinfo {author} {\bibfnamefont {W.}~\bibnamefont
  {Casteels}}\ and\ \bibinfo {author} {\bibfnamefont {C.}~\bibnamefont
  {Ciuti}},\ }\bibfield  {title} {\bibinfo {title} {Quantum entanglement in the
  spatial-symmetry-breaking phase transition of a driven-dissipative
  bose-hubbard dimer},\ }\href {https://doi.org/10.1103/PhysRevA.95.013812}
  {\bibfield  {journal} {\bibinfo  {journal} {Phys. Rev. A}\ }\textbf {\bibinfo
  {volume} {95}},\ \bibinfo {pages} {013812} (\bibinfo {year}
  {2017})}\BibitemShut {NoStop}%
\bibitem [{\citenamefont {Casteels}\ \emph {et~al.}(2017)\citenamefont
  {Casteels}, \citenamefont {Fazio},\ and\ \citenamefont
  {Ciuti}}]{Casteels17_1stOrderDPT}%
  \BibitemOpen
  \bibfield  {author} {\bibinfo {author} {\bibfnamefont {W.}~\bibnamefont
  {Casteels}}, \bibinfo {author} {\bibfnamefont {R.}~\bibnamefont {Fazio}},\
  and\ \bibinfo {author} {\bibfnamefont {C.}~\bibnamefont {Ciuti}},\ }\bibfield
   {title} {\bibinfo {title} {Critical dynamical properties of a first-order
  dissipative phase transition},\ }\href
  {https://doi.org/10.1103/PhysRevA.95.012128} {\bibfield  {journal} {\bibinfo
  {journal} {Phys. Rev. A}\ }\textbf {\bibinfo {volume} {95}},\ \bibinfo
  {pages} {012128} (\bibinfo {year} {2017})}\BibitemShut {NoStop}%
\bibitem [{\citenamefont {Foss-Feig}\ \emph {et~al.}(2017)\citenamefont
  {Foss-Feig}, \citenamefont {Niroula}, \citenamefont {Young}, \citenamefont
  {Hafezi}, \citenamefont {Gorshkov}, \citenamefont {Wilson},\ and\
  \citenamefont {Maghrebi}}]{Foss-Feig17}%
  \BibitemOpen
  \bibfield  {author} {\bibinfo {author} {\bibfnamefont {M.}~\bibnamefont
  {Foss-Feig}}, \bibinfo {author} {\bibfnamefont {P.}~\bibnamefont {Niroula}},
  \bibinfo {author} {\bibfnamefont {J.~T.}\ \bibnamefont {Young}}, \bibinfo
  {author} {\bibfnamefont {M.}~\bibnamefont {Hafezi}}, \bibinfo {author}
  {\bibfnamefont {A.~V.}\ \bibnamefont {Gorshkov}}, \bibinfo {author}
  {\bibfnamefont {R.~M.}\ \bibnamefont {Wilson}},\ and\ \bibinfo {author}
  {\bibfnamefont {M.~F.}\ \bibnamefont {Maghrebi}},\ }\bibfield  {title}
  {\bibinfo {title} {Emergent equilibrium in many-body optical bistability},\
  }\href {https://doi.org/10.1103/PhysRevA.95.043826} {\bibfield  {journal}
  {\bibinfo  {journal} {Phys. Rev. A}\ }\textbf {\bibinfo {volume} {95}},\
  \bibinfo {pages} {043826} (\bibinfo {year} {2017})}\BibitemShut {NoStop}%
\bibitem [{\citenamefont {Biondi}\ \emph {et~al.}(2017)\citenamefont {Biondi},
  \citenamefont {Blatter}, \citenamefont {T\"ureci},\ and\ \citenamefont
  {Schmidt}}]{Biondi17}%
  \BibitemOpen
  \bibfield  {author} {\bibinfo {author} {\bibfnamefont {M.}~\bibnamefont
  {Biondi}}, \bibinfo {author} {\bibfnamefont {G.}~\bibnamefont {Blatter}},
  \bibinfo {author} {\bibfnamefont {H.~E.}\ \bibnamefont {T\"ureci}},\ and\
  \bibinfo {author} {\bibfnamefont {S.}~\bibnamefont {Schmidt}},\ }\bibfield
  {title} {\bibinfo {title} {Nonequilibrium gas-liquid transition in the
  driven-dissipative photonic lattice},\ }\href
  {https://doi.org/10.1103/PhysRevA.96.043809} {\bibfield  {journal} {\bibinfo
  {journal} {Phys. Rev. A}\ }\textbf {\bibinfo {volume} {96}},\ \bibinfo
  {pages} {043809} (\bibinfo {year} {2017})}\BibitemShut {NoStop}%
\bibitem [{\citenamefont {Biella}\ \emph {et~al.}(2017)\citenamefont {Biella},
  \citenamefont {Storme}, \citenamefont {Lebreuilly}, \citenamefont {Rossini},
  \citenamefont {Fazio}, \citenamefont {Carusotto},\ and\ \citenamefont
  {Ciuti}}]{Biella17}%
  \BibitemOpen
  \bibfield  {author} {\bibinfo {author} {\bibfnamefont {A.}~\bibnamefont
  {Biella}}, \bibinfo {author} {\bibfnamefont {F.}~\bibnamefont {Storme}},
  \bibinfo {author} {\bibfnamefont {J.}~\bibnamefont {Lebreuilly}}, \bibinfo
  {author} {\bibfnamefont {D.}~\bibnamefont {Rossini}}, \bibinfo {author}
  {\bibfnamefont {R.}~\bibnamefont {Fazio}}, \bibinfo {author} {\bibfnamefont
  {I.}~\bibnamefont {Carusotto}},\ and\ \bibinfo {author} {\bibfnamefont
  {C.}~\bibnamefont {Ciuti}},\ }\bibfield  {title} {\bibinfo {title} {Phase
  diagram of incoherently driven strongly correlated photonic lattices},\
  }\href {https://doi.org/10.1103/PhysRevA.96.023839} {\bibfield  {journal}
  {\bibinfo  {journal} {Phys. Rev. A}\ }\textbf {\bibinfo {volume} {96}},\
  \bibinfo {pages} {023839} (\bibinfo {year} {2017})}\BibitemShut {NoStop}%
\bibitem [{\citenamefont {Savona}(2017)}]{VincenzoMF}%
  \BibitemOpen
  \bibfield  {author} {\bibinfo {author} {\bibfnamefont {V.}~\bibnamefont
  {Savona}},\ }\bibfield  {title} {\bibinfo {title} {Spontaneous symmetry
  breaking in a quadratically driven nonlinear photonic lattice},\ }\href
  {https://doi.org/10.1103/PhysRevA.96.033826} {\bibfield  {journal} {\bibinfo
  {journal} {Phys. Rev. A}\ }\textbf {\bibinfo {volume} {96}},\ \bibinfo
  {pages} {033826} (\bibinfo {year} {2017})}\BibitemShut {NoStop}%
\bibitem [{\citenamefont {Vicentini}\ \emph {et~al.}(2018)\citenamefont
  {Vicentini}, \citenamefont {Minganti}, \citenamefont {Rota}, \citenamefont
  {Orso},\ and\ \citenamefont {Ciuti}}]{Vicentini18}%
  \BibitemOpen
  \bibfield  {author} {\bibinfo {author} {\bibfnamefont {F.}~\bibnamefont
  {Vicentini}}, \bibinfo {author} {\bibfnamefont {F.}~\bibnamefont {Minganti}},
  \bibinfo {author} {\bibfnamefont {R.}~\bibnamefont {Rota}}, \bibinfo {author}
  {\bibfnamefont {G.}~\bibnamefont {Orso}},\ and\ \bibinfo {author}
  {\bibfnamefont {C.}~\bibnamefont {Ciuti}},\ }\bibfield  {title} {\bibinfo
  {title} {Critical slowing down in driven-dissipative bose-hubbard lattices},\
  }\href {https://doi.org/10.1103/PhysRevA.97.013853} {\bibfield  {journal}
  {\bibinfo  {journal} {Phys. Rev. A}\ }\textbf {\bibinfo {volume} {97}},\
  \bibinfo {pages} {013853} (\bibinfo {year} {2018})}\BibitemShut {NoStop}%
\bibitem [{\citenamefont {Rota}\ \emph {et~al.}(2019)\citenamefont {Rota},
  \citenamefont {Minganti}, \citenamefont {Ciuti},\ and\ \citenamefont
  {Savona}}]{PRLspinmodel}%
  \BibitemOpen
  \bibfield  {author} {\bibinfo {author} {\bibfnamefont {R.}~\bibnamefont
  {Rota}}, \bibinfo {author} {\bibfnamefont {F.}~\bibnamefont {Minganti}},
  \bibinfo {author} {\bibfnamefont {C.}~\bibnamefont {Ciuti}},\ and\ \bibinfo
  {author} {\bibfnamefont {V.}~\bibnamefont {Savona}},\ }\bibfield  {title}
  {\bibinfo {title} {Quantum critical regime in a quadratically driven
  nonlinear photonic lattice},\ }\href
  {https://doi.org/10.1103/PhysRevLett.122.110405} {\bibfield  {journal}
  {\bibinfo  {journal} {Phys. Rev. Lett.}\ }\textbf {\bibinfo {volume} {122}},\
  \bibinfo {pages} {110405} (\bibinfo {year} {2019})}\BibitemShut {NoStop}%
\bibitem [{\citenamefont {Sieberer}\ \emph {et~al.}(2013)\citenamefont
  {Sieberer}, \citenamefont {Huber}, \citenamefont {Altman},\ and\
  \citenamefont {Diehl}}]{Sieberer13}%
  \BibitemOpen
  \bibfield  {author} {\bibinfo {author} {\bibfnamefont {L.~M.}\ \bibnamefont
  {Sieberer}}, \bibinfo {author} {\bibfnamefont {S.~D.}\ \bibnamefont {Huber}},
  \bibinfo {author} {\bibfnamefont {E.}~\bibnamefont {Altman}},\ and\ \bibinfo
  {author} {\bibfnamefont {S.}~\bibnamefont {Diehl}},\ }\bibfield  {title}
  {\bibinfo {title} {Dynamical critical phenomena in driven-dissipative
  systems},\ }\href {https://doi.org/10.1103/PhysRevLett.110.195301} {\bibfield
   {journal} {\bibinfo  {journal} {Phys. Rev. Lett.}\ }\textbf {\bibinfo
  {volume} {110}},\ \bibinfo {pages} {195301} (\bibinfo {year}
  {2013})}\BibitemShut {NoStop}%
\bibitem [{\citenamefont {Sieberer}\ \emph {et~al.}(2014)\citenamefont
  {Sieberer}, \citenamefont {Huber}, \citenamefont {Altman},\ and\
  \citenamefont {Diehl}}]{Sieberer14}%
  \BibitemOpen
  \bibfield  {author} {\bibinfo {author} {\bibfnamefont {L.~M.}\ \bibnamefont
  {Sieberer}}, \bibinfo {author} {\bibfnamefont {S.~D.}\ \bibnamefont {Huber}},
  \bibinfo {author} {\bibfnamefont {E.}~\bibnamefont {Altman}},\ and\ \bibinfo
  {author} {\bibfnamefont {S.}~\bibnamefont {Diehl}},\ }\bibfield  {title}
  {\bibinfo {title} {Nonequilibrium functional renormalization for
  driven-dissipative bose-einstein condensation},\ }\href
  {https://doi.org/10.1103/PhysRevB.89.134310} {\bibfield  {journal} {\bibinfo
  {journal} {Phys. Rev. B}\ }\textbf {\bibinfo {volume} {89}},\ \bibinfo
  {pages} {134310} (\bibinfo {year} {2014})}\BibitemShut {NoStop}%
\bibitem [{\citenamefont {Altman}\ \emph {et~al.}(2015)\citenamefont {Altman},
  \citenamefont {Sieberer}, \citenamefont {Chen}, \citenamefont {Diehl},\ and\
  \citenamefont {Toner}}]{Altman15}%
  \BibitemOpen
  \bibfield  {author} {\bibinfo {author} {\bibfnamefont {E.}~\bibnamefont
  {Altman}}, \bibinfo {author} {\bibfnamefont {L.~M.}\ \bibnamefont
  {Sieberer}}, \bibinfo {author} {\bibfnamefont {L.}~\bibnamefont {Chen}},
  \bibinfo {author} {\bibfnamefont {S.}~\bibnamefont {Diehl}},\ and\ \bibinfo
  {author} {\bibfnamefont {J.}~\bibnamefont {Toner}},\ }\bibfield  {title}
  {\bibinfo {title} {Two-dimensional superfluidity of exciton polaritons
  requires strong anisotropy},\ }\href
  {https://doi.org/10.1103/PhysRevX.5.011017} {\bibfield  {journal} {\bibinfo
  {journal} {Phys. Rev. X}\ }\textbf {\bibinfo {volume} {5}},\ \bibinfo {pages}
  {011017} (\bibinfo {year} {2015})}\BibitemShut {NoStop}%
\bibitem [{\citenamefont {Lee}\ \emph {et~al.}(2011)\citenamefont {Lee},
  \citenamefont {H\"affner},\ and\ \citenamefont {Cross}}]{Lee11}%
  \BibitemOpen
  \bibfield  {author} {\bibinfo {author} {\bibfnamefont {T.~E.}\ \bibnamefont
  {Lee}}, \bibinfo {author} {\bibfnamefont {H.}~\bibnamefont {H\"affner}},\
  and\ \bibinfo {author} {\bibfnamefont {M.~C.}\ \bibnamefont {Cross}},\
  }\bibfield  {title} {\bibinfo {title} {Antiferromagnetic phase transition in
  a nonequilibrium lattice of rydberg atoms},\ }\href
  {https://doi.org/10.1103/PhysRevA.84.031402} {\bibfield  {journal} {\bibinfo
  {journal} {Phys. Rev. A}\ }\textbf {\bibinfo {volume} {84}},\ \bibinfo
  {pages} {031402(R)} (\bibinfo {year} {2011})}\BibitemShut {NoStop}%
\bibitem [{\citenamefont {Lee}\ \emph {et~al.}(2013)\citenamefont {Lee},
  \citenamefont {Gopalakrishnan},\ and\ \citenamefont {Lukin}}]{Lee13}%
  \BibitemOpen
  \bibfield  {author} {\bibinfo {author} {\bibfnamefont {T.~E.}\ \bibnamefont
  {Lee}}, \bibinfo {author} {\bibfnamefont {S.}~\bibnamefont
  {Gopalakrishnan}},\ and\ \bibinfo {author} {\bibfnamefont {M.~D.}\
  \bibnamefont {Lukin}},\ }\bibfield  {title} {\bibinfo {title} {Unconventional
  magnetism via optical pumping of interacting spin systems},\ }\href
  {https://doi.org/10.1103/PhysRevLett.110.257204} {\bibfield  {journal}
  {\bibinfo  {journal} {Phys. Rev. Lett.}\ }\textbf {\bibinfo {volume} {110}},\
  \bibinfo {pages} {257204} (\bibinfo {year} {2013})}\BibitemShut {NoStop}%
\bibitem [{\citenamefont {Chan}\ \emph {et~al.}(2015)\citenamefont {Chan},
  \citenamefont {Lee},\ and\ \citenamefont {Gopalakrishnan}}]{Chan15}%
  \BibitemOpen
  \bibfield  {author} {\bibinfo {author} {\bibfnamefont {C.-K.}\ \bibnamefont
  {Chan}}, \bibinfo {author} {\bibfnamefont {T.~E.}\ \bibnamefont {Lee}},\ and\
  \bibinfo {author} {\bibfnamefont {S.}~\bibnamefont {Gopalakrishnan}},\
  }\bibfield  {title} {\bibinfo {title} {Limit-cycle phase in
  driven-dissipative spin systems},\ }\href
  {https://doi.org/10.1103/PhysRevA.91.051601} {\bibfield  {journal} {\bibinfo
  {journal} {Phys. Rev. A}\ }\textbf {\bibinfo {volume} {91}},\ \bibinfo
  {pages} {051601(R)} (\bibinfo {year} {2015})}\BibitemShut {NoStop}%
\bibitem [{\citenamefont {Jin}\ \emph {et~al.}(2016)\citenamefont {Jin},
  \citenamefont {Biella}, \citenamefont {Viyuela}, \citenamefont {Mazza},
  \citenamefont {Keeling}, \citenamefont {Fazio},\ and\ \citenamefont
  {Rossini}}]{Jin16}%
  \BibitemOpen
  \bibfield  {author} {\bibinfo {author} {\bibfnamefont {J.}~\bibnamefont
  {Jin}}, \bibinfo {author} {\bibfnamefont {A.}~\bibnamefont {Biella}},
  \bibinfo {author} {\bibfnamefont {O.}~\bibnamefont {Viyuela}}, \bibinfo
  {author} {\bibfnamefont {L.}~\bibnamefont {Mazza}}, \bibinfo {author}
  {\bibfnamefont {J.}~\bibnamefont {Keeling}}, \bibinfo {author} {\bibfnamefont
  {R.}~\bibnamefont {Fazio}},\ and\ \bibinfo {author} {\bibfnamefont
  {D.}~\bibnamefont {Rossini}},\ }\bibfield  {title} {\bibinfo {title} {Cluster
  mean-field approach to the steady-state phase diagram of dissipative spin
  systems},\ }\href {https://doi.org/10.1103/PhysRevX.6.031011} {\bibfield
  {journal} {\bibinfo  {journal} {Phys. Rev. X}\ }\textbf {\bibinfo {volume}
  {6}},\ \bibinfo {pages} {031011} (\bibinfo {year} {2016})}\BibitemShut
  {NoStop}%
\bibitem [{\citenamefont {Maghrebi}\ and\ \citenamefont
  {Gorshkov}(2016)}]{Maghrebi16}%
  \BibitemOpen
  \bibfield  {author} {\bibinfo {author} {\bibfnamefont {M.~F.}\ \bibnamefont
  {Maghrebi}}\ and\ \bibinfo {author} {\bibfnamefont {A.~V.}\ \bibnamefont
  {Gorshkov}},\ }\bibfield  {title} {\bibinfo {title} {Nonequilibrium many-body
  steady states via keldysh formalism},\ }\href
  {https://doi.org/10.1103/PhysRevB.93.014307} {\bibfield  {journal} {\bibinfo
  {journal} {Phys. Rev. B}\ }\textbf {\bibinfo {volume} {93}},\ \bibinfo
  {pages} {014307} (\bibinfo {year} {2016})}\BibitemShut {NoStop}%
\bibitem [{\citenamefont {Rota}\ \emph {et~al.}(2017)\citenamefont {Rota},
  \citenamefont {Storme}, \citenamefont {Bartolo}, \citenamefont {Fazio},\ and\
  \citenamefont {Ciuti}}]{Rota17}%
  \BibitemOpen
  \bibfield  {author} {\bibinfo {author} {\bibfnamefont {R.}~\bibnamefont
  {Rota}}, \bibinfo {author} {\bibfnamefont {F.}~\bibnamefont {Storme}},
  \bibinfo {author} {\bibfnamefont {N.}~\bibnamefont {Bartolo}}, \bibinfo
  {author} {\bibfnamefont {R.}~\bibnamefont {Fazio}},\ and\ \bibinfo {author}
  {\bibfnamefont {C.}~\bibnamefont {Ciuti}},\ }\bibfield  {title} {\bibinfo
  {title} {Critical behavior of dissipative two-dimensional spin lattices},\
  }\href {https://doi.org/10.1103/PhysRevB.95.134431} {\bibfield  {journal}
  {\bibinfo  {journal} {Phys. Rev. B}\ }\textbf {\bibinfo {volume} {95}},\
  \bibinfo {pages} {134431} (\bibinfo {year} {2017})}\BibitemShut {NoStop}%
\bibitem [{\citenamefont {Overbeck}\ \emph {et~al.}(2017)\citenamefont
  {Overbeck}, \citenamefont {Maghrebi}, \citenamefont {Gorshkov},\ and\
  \citenamefont {Weimer}}]{Overbeck17}%
  \BibitemOpen
  \bibfield  {author} {\bibinfo {author} {\bibfnamefont {V.~R.}\ \bibnamefont
  {Overbeck}}, \bibinfo {author} {\bibfnamefont {M.~F.}\ \bibnamefont
  {Maghrebi}}, \bibinfo {author} {\bibfnamefont {A.~V.}\ \bibnamefont
  {Gorshkov}},\ and\ \bibinfo {author} {\bibfnamefont {H.}~\bibnamefont
  {Weimer}},\ }\bibfield  {title} {\bibinfo {title} {Multicritical behavior in
  dissipative ising models},\ }\href
  {https://doi.org/10.1103/PhysRevA.95.042133} {\bibfield  {journal} {\bibinfo
  {journal} {Phys. Rev. A}\ }\textbf {\bibinfo {volume} {95}},\ \bibinfo
  {pages} {042133} (\bibinfo {year} {2017})}\BibitemShut {NoStop}%
\bibitem [{\citenamefont {Rota}\ \emph {et~al.}(2018)\citenamefont {Rota},
  \citenamefont {Minganti}, \citenamefont {Biella},\ and\ \citenamefont
  {Ciuti}}]{XYZdynamics}%
  \BibitemOpen
  \bibfield  {author} {\bibinfo {author} {\bibfnamefont {R.}~\bibnamefont
  {Rota}}, \bibinfo {author} {\bibfnamefont {F.}~\bibnamefont {Minganti}},
  \bibinfo {author} {\bibfnamefont {A.}~\bibnamefont {Biella}},\ and\ \bibinfo
  {author} {\bibfnamefont {C.}~\bibnamefont {Ciuti}},\ }\bibfield  {title}
  {\bibinfo {title} {Dynamical properties of dissipative {XYZ} heisenberg
  lattices},\ }\href {https://doi.org/10.1088/1367-2630/aab703} {\bibfield
  {journal} {\bibinfo  {journal} {New Journal of Physics}\ }\textbf {\bibinfo
  {volume} {20}},\ \bibinfo {pages} {045003} (\bibinfo {year}
  {2018})}\BibitemShut {NoStop}%
\bibitem [{\citenamefont {Roscher}\ \emph {et~al.}(2018)\citenamefont
  {Roscher}, \citenamefont {Diehl},\ and\ \citenamefont
  {Buchhold}}]{Roscher18}%
  \BibitemOpen
  \bibfield  {author} {\bibinfo {author} {\bibfnamefont {D.}~\bibnamefont
  {Roscher}}, \bibinfo {author} {\bibfnamefont {S.}~\bibnamefont {Diehl}},\
  and\ \bibinfo {author} {\bibfnamefont {M.}~\bibnamefont {Buchhold}},\
  }\bibfield  {title} {\bibinfo {title} {Phenomenology of first-order
  dark-state phase transitions},\ }\href
  {https://doi.org/10.1103/PhysRevA.98.062117} {\bibfield  {journal} {\bibinfo
  {journal} {Phys. Rev. A}\ }\textbf {\bibinfo {volume} {98}},\ \bibinfo
  {pages} {062117} (\bibinfo {year} {2018})}\BibitemShut {NoStop}%
\bibitem [{\citenamefont {Fink}\ \emph {et~al.}(2017)\citenamefont {Fink},
  \citenamefont {Dombi}, \citenamefont {Vukics}, \citenamefont {Wallraff},\
  and\ \citenamefont {Domokos}}]{Fink17}%
  \BibitemOpen
  \bibfield  {author} {\bibinfo {author} {\bibfnamefont {J.~M.}\ \bibnamefont
  {Fink}}, \bibinfo {author} {\bibfnamefont {A.}~\bibnamefont {Dombi}},
  \bibinfo {author} {\bibfnamefont {A.}~\bibnamefont {Vukics}}, \bibinfo
  {author} {\bibfnamefont {A.}~\bibnamefont {Wallraff}},\ and\ \bibinfo
  {author} {\bibfnamefont {P.}~\bibnamefont {Domokos}},\ }\bibfield  {title}
  {\bibinfo {title} {Observation of the photon-blockade breakdown phase
  transition},\ }\href {https://doi.org/10.1103/PhysRevX.7.011012} {\bibfield
  {journal} {\bibinfo  {journal} {Phys. Rev. X}\ }\textbf {\bibinfo {volume}
  {7}},\ \bibinfo {pages} {011012} (\bibinfo {year} {2017})}\BibitemShut
  {NoStop}%
\bibitem [{\citenamefont {Rodriguez}\ \emph {et~al.}(2017)\citenamefont
  {Rodriguez}, \citenamefont {Casteels}, \citenamefont {Storme}, \citenamefont
  {Carlon~Zambon}, \citenamefont {Sagnes}, \citenamefont {Le~Gratiet},
  \citenamefont {Galopin}, \citenamefont {Lema\^{\i}tre}, \citenamefont {Amo},
  \citenamefont {Ciuti},\ and\ \citenamefont {Bloch}}]{Rodriguez17}%
  \BibitemOpen
  \bibfield  {author} {\bibinfo {author} {\bibfnamefont {S.~R.~K.}\
  \bibnamefont {Rodriguez}}, \bibinfo {author} {\bibfnamefont {W.}~\bibnamefont
  {Casteels}}, \bibinfo {author} {\bibfnamefont {F.}~\bibnamefont {Storme}},
  \bibinfo {author} {\bibfnamefont {N.}~\bibnamefont {Carlon~Zambon}}, \bibinfo
  {author} {\bibfnamefont {I.}~\bibnamefont {Sagnes}}, \bibinfo {author}
  {\bibfnamefont {L.}~\bibnamefont {Le~Gratiet}}, \bibinfo {author}
  {\bibfnamefont {E.}~\bibnamefont {Galopin}}, \bibinfo {author} {\bibfnamefont
  {A.}~\bibnamefont {Lema\^{\i}tre}}, \bibinfo {author} {\bibfnamefont
  {A.}~\bibnamefont {Amo}}, \bibinfo {author} {\bibfnamefont {C.}~\bibnamefont
  {Ciuti}},\ and\ \bibinfo {author} {\bibfnamefont {J.}~\bibnamefont {Bloch}},\
  }\bibfield  {title} {\bibinfo {title} {Probing a dissipative phase transition
  via dynamical optical hysteresis},\ }\href
  {https://doi.org/10.1103/PhysRevLett.118.247402} {\bibfield  {journal}
  {\bibinfo  {journal} {Phys. Rev. Lett.}\ }\textbf {\bibinfo {volume} {118}},\
  \bibinfo {pages} {247402} (\bibinfo {year} {2017})}\BibitemShut {NoStop}%
\bibitem [{\citenamefont {Fink}\ \emph {et~al.}(2018)\citenamefont {Fink},
  \citenamefont {Schade}, \citenamefont {H{\"o}fling}, \citenamefont
  {Schneider},\ and\ \citenamefont {Imamoglu}}]{Fink2018}%
  \BibitemOpen
  \bibfield  {author} {\bibinfo {author} {\bibfnamefont {T.}~\bibnamefont
  {Fink}}, \bibinfo {author} {\bibfnamefont {A.}~\bibnamefont {Schade}},
  \bibinfo {author} {\bibfnamefont {S.}~\bibnamefont {H{\"o}fling}}, \bibinfo
  {author} {\bibfnamefont {C.}~\bibnamefont {Schneider}},\ and\ \bibinfo
  {author} {\bibfnamefont {A.}~\bibnamefont {Imamoglu}},\ }\bibfield  {title}
  {\bibinfo {title} {Signatures of a dissipative phase transition in photon
  correlation measurements},\ }\href
  {https://doi.org/10.1038/s41567-017-0020-9} {\bibfield  {journal} {\bibinfo
  {journal} {Nature Physics}\ }\textbf {\bibinfo {volume} {14}},\ \bibinfo
  {pages} {365} (\bibinfo {year} {2018})}\BibitemShut {NoStop}%
\bibitem [{\citenamefont {Diehl}\ \emph {et~al.}(2010)\citenamefont {Diehl},
  \citenamefont {Tomadin}, \citenamefont {Micheli}, \citenamefont {Fazio},\
  and\ \citenamefont {Zoller}}]{Diehl10}%
  \BibitemOpen
  \bibfield  {author} {\bibinfo {author} {\bibfnamefont {S.}~\bibnamefont
  {Diehl}}, \bibinfo {author} {\bibfnamefont {A.}~\bibnamefont {Tomadin}},
  \bibinfo {author} {\bibfnamefont {A.}~\bibnamefont {Micheli}}, \bibinfo
  {author} {\bibfnamefont {R.}~\bibnamefont {Fazio}},\ and\ \bibinfo {author}
  {\bibfnamefont {P.}~\bibnamefont {Zoller}},\ }\bibfield  {title} {\bibinfo
  {title} {Dynamical phase transitions and instabilities in open atomic
  many-body systems},\ }\href {https://doi.org/10.1103/PhysRevLett.105.015702}
  {\bibfield  {journal} {\bibinfo  {journal} {Phys. Rev. Lett.}\ }\textbf
  {\bibinfo {volume} {105}},\ \bibinfo {pages} {015702} (\bibinfo {year}
  {2010})}\BibitemShut {NoStop}%
\bibitem [{\citenamefont {Dalla~Torre}\ \emph {et~al.}(2010)\citenamefont
  {Dalla~Torre}, \citenamefont {Demler}, \citenamefont {Giamarchi},\ and\
  \citenamefont {Altman}}]{DallaTorrethermalno}%
  \BibitemOpen
  \bibfield  {author} {\bibinfo {author} {\bibfnamefont {E.~G.}\ \bibnamefont
  {Dalla~Torre}}, \bibinfo {author} {\bibfnamefont {E.}~\bibnamefont {Demler}},
  \bibinfo {author} {\bibfnamefont {T.}~\bibnamefont {Giamarchi}},\ and\
  \bibinfo {author} {\bibfnamefont {E.}~\bibnamefont {Altman}},\ }\bibfield
  {title} {\bibinfo {title} {Quantum critical states and phase transitions in
  the presence of non-equilibrium noise},\ }\href
  {https://doi.org/10.1038/nphys1754} {\bibfield  {journal} {\bibinfo
  {journal} {Nature Physics}\ }\textbf {\bibinfo {volume} {6}},\ \bibinfo
  {pages} {806} (\bibinfo {year} {2010})}\BibitemShut {NoStop}%
\bibitem [{\citenamefont {Dalla~Torre}\ \emph {et~al.}(2012)\citenamefont
  {Dalla~Torre}, \citenamefont {Demler}, \citenamefont {Giamarchi},\ and\
  \citenamefont {Altman}}]{DallaTorrethermalyes}%
  \BibitemOpen
  \bibfield  {author} {\bibinfo {author} {\bibfnamefont {E.~G.}\ \bibnamefont
  {Dalla~Torre}}, \bibinfo {author} {\bibfnamefont {E.}~\bibnamefont {Demler}},
  \bibinfo {author} {\bibfnamefont {T.}~\bibnamefont {Giamarchi}},\ and\
  \bibinfo {author} {\bibfnamefont {E.}~\bibnamefont {Altman}},\ }\bibfield
  {title} {\bibinfo {title} {Dynamics and universality in noise-driven
  dissipative systems},\ }\href {https://doi.org/10.1103/PhysRevB.85.184302}
  {\bibfield  {journal} {\bibinfo  {journal} {Phys. Rev. B}\ }\textbf {\bibinfo
  {volume} {85}},\ \bibinfo {pages} {184302} (\bibinfo {year}
  {2012})}\BibitemShut {NoStop}%
\bibitem [{\citenamefont {T\"auber}\ and\ \citenamefont
  {Diehl}(2014)}]{Tauber14}%
  \BibitemOpen
  \bibfield  {author} {\bibinfo {author} {\bibfnamefont {U.~C.}\ \bibnamefont
  {T\"auber}}\ and\ \bibinfo {author} {\bibfnamefont {S.}~\bibnamefont
  {Diehl}},\ }\bibfield  {title} {\bibinfo {title} {Perturbative
  field-theoretical renormalization group approach to driven-dissipative
  bose-einstein criticality},\ }\href
  {https://doi.org/10.1103/PhysRevX.4.021010} {\bibfield  {journal} {\bibinfo
  {journal} {Phys. Rev. X}\ }\textbf {\bibinfo {volume} {4}},\ \bibinfo {pages}
  {021010} (\bibinfo {year} {2014})}\BibitemShut {NoStop}%
\bibitem [{\citenamefont {Marino}\ and\ \citenamefont
  {Diehl}(2016)}]{Diehlnewuniversalityclass}%
  \BibitemOpen
  \bibfield  {author} {\bibinfo {author} {\bibfnamefont {J.}~\bibnamefont
  {Marino}}\ and\ \bibinfo {author} {\bibfnamefont {S.}~\bibnamefont {Diehl}},\
  }\bibfield  {title} {\bibinfo {title} {Driven markovian quantum
  criticality},\ }\href {https://doi.org/10.1103/PhysRevLett.116.070407}
  {\bibfield  {journal} {\bibinfo  {journal} {Phys. Rev. Lett.}\ }\textbf
  {\bibinfo {volume} {116}},\ \bibinfo {pages} {070407} (\bibinfo {year}
  {2016})}\BibitemShut {NoStop}%
\bibitem [{\citenamefont {Mitra}\ \emph {et~al.}(2006)\citenamefont {Mitra},
  \citenamefont {Takei}, \citenamefont {Kim},\ and\ \citenamefont
  {Millis}}]{Mitra06}%
  \BibitemOpen
  \bibfield  {author} {\bibinfo {author} {\bibfnamefont {A.}~\bibnamefont
  {Mitra}}, \bibinfo {author} {\bibfnamefont {S.}~\bibnamefont {Takei}},
  \bibinfo {author} {\bibfnamefont {Y.~B.}\ \bibnamefont {Kim}},\ and\ \bibinfo
  {author} {\bibfnamefont {A.~J.}\ \bibnamefont {Millis}},\ }\bibfield  {title}
  {\bibinfo {title} {Nonequilibrium quantum criticality in open electronic
  systems},\ }\href {https://doi.org/10.1103/PhysRevLett.97.236808} {\bibfield
  {journal} {\bibinfo  {journal} {Phys. Rev. Lett.}\ }\textbf {\bibinfo
  {volume} {97}},\ \bibinfo {pages} {236808} (\bibinfo {year}
  {2006})}\BibitemShut {NoStop}%
\bibitem [{\citenamefont {Torre}\ \emph {et~al.}(2013)\citenamefont {Torre},
  \citenamefont {Diehl}, \citenamefont {Lukin}, \citenamefont {Sachdev},\ and\
  \citenamefont {Strack}}]{DallaTorre13}%
  \BibitemOpen
  \bibfield  {author} {\bibinfo {author} {\bibfnamefont {E.~G.~D.}\
  \bibnamefont {Torre}}, \bibinfo {author} {\bibfnamefont {S.}~\bibnamefont
  {Diehl}}, \bibinfo {author} {\bibfnamefont {M.~D.}\ \bibnamefont {Lukin}},
  \bibinfo {author} {\bibfnamefont {S.}~\bibnamefont {Sachdev}},\ and\ \bibinfo
  {author} {\bibfnamefont {P.}~\bibnamefont {Strack}},\ }\bibfield  {title}
  {\bibinfo {title} {Keldysh approach for nonequilibrium phase transitions in
  quantum optics: Beyond the dicke model in optical cavities},\ }\href
  {https://doi.org/10.1103/PhysRevA.87.023831} {\bibfield  {journal} {\bibinfo
  {journal} {Phys. Rev. A}\ }\textbf {\bibinfo {volume} {87}},\ \bibinfo
  {pages} {023831} (\bibinfo {year} {2013})}\BibitemShut {NoStop}%
\bibitem [{\citenamefont {Leghtas}\ \emph {et~al.}(2015)\citenamefont
  {Leghtas}, \citenamefont {Touzard}, \citenamefont {Pop}, \citenamefont {Kou},
  \citenamefont {Vlastakis}, \citenamefont {Petrenko}, \citenamefont {Sliwa},
  \citenamefont {Narla}, \citenamefont {Shankar}, \citenamefont {Hatridge},
  \citenamefont {Reagor}, \citenamefont {Frunzio}, \citenamefont {Schoelkopf},
  \citenamefont {Mirrahimi},\ and\ \citenamefont {Devoret}}]{Leghtas15}%
  \BibitemOpen
  \bibfield  {author} {\bibinfo {author} {\bibfnamefont {Z.}~\bibnamefont
  {Leghtas}}, \bibinfo {author} {\bibfnamefont {S.}~\bibnamefont {Touzard}},
  \bibinfo {author} {\bibfnamefont {I.~M.}\ \bibnamefont {Pop}}, \bibinfo
  {author} {\bibfnamefont {A.}~\bibnamefont {Kou}}, \bibinfo {author}
  {\bibfnamefont {B.}~\bibnamefont {Vlastakis}}, \bibinfo {author}
  {\bibfnamefont {A.}~\bibnamefont {Petrenko}}, \bibinfo {author}
  {\bibfnamefont {K.~M.}\ \bibnamefont {Sliwa}}, \bibinfo {author}
  {\bibfnamefont {A.}~\bibnamefont {Narla}}, \bibinfo {author} {\bibfnamefont
  {S.}~\bibnamefont {Shankar}}, \bibinfo {author} {\bibfnamefont {M.~J.}\
  \bibnamefont {Hatridge}}, \bibinfo {author} {\bibfnamefont {M.}~\bibnamefont
  {Reagor}}, \bibinfo {author} {\bibfnamefont {L.}~\bibnamefont {Frunzio}},
  \bibinfo {author} {\bibfnamefont {R.~J.}\ \bibnamefont {Schoelkopf}},
  \bibinfo {author} {\bibfnamefont {M.}~\bibnamefont {Mirrahimi}},\ and\
  \bibinfo {author} {\bibfnamefont {M.~H.}\ \bibnamefont {Devoret}},\
  }\bibfield  {title} {\bibinfo {title} {Confining the state of light to a
  quantum manifold by engineered two-photon loss},\ }\href
  {https://doi.org/10.1126/science.aaa2085} {\bibfield  {journal} {\bibinfo
  {journal} {Science}\ }\textbf {\bibinfo {volume} {347}},\ \bibinfo {pages}
  {853} (\bibinfo {year} {2015})}\BibitemShut {NoStop}%
\bibitem [{\citenamefont {Wang}\ \emph {et~al.}(2016)\citenamefont {Wang},
  \citenamefont {Gao}, \citenamefont {Reinhold}, \citenamefont {Heeres},
  \citenamefont {Ofek}, \citenamefont {Chou}, \citenamefont {Axline},
  \citenamefont {Reagor}, \citenamefont {Blumoff}, \citenamefont {Sliwa},
  \citenamefont {Frunzio}, \citenamefont {Girvin}, \citenamefont {Jiang},
  \citenamefont {Mirrahimi}, \citenamefont {Devoret},\ and\ \citenamefont
  {Schoelkopf}}]{Wang16}%
  \BibitemOpen
  \bibfield  {author} {\bibinfo {author} {\bibfnamefont {C.}~\bibnamefont
  {Wang}}, \bibinfo {author} {\bibfnamefont {Y.~Y.}\ \bibnamefont {Gao}},
  \bibinfo {author} {\bibfnamefont {P.}~\bibnamefont {Reinhold}}, \bibinfo
  {author} {\bibfnamefont {R.~W.}\ \bibnamefont {Heeres}}, \bibinfo {author}
  {\bibfnamefont {N.}~\bibnamefont {Ofek}}, \bibinfo {author} {\bibfnamefont
  {K.}~\bibnamefont {Chou}}, \bibinfo {author} {\bibfnamefont {C.}~\bibnamefont
  {Axline}}, \bibinfo {author} {\bibfnamefont {M.}~\bibnamefont {Reagor}},
  \bibinfo {author} {\bibfnamefont {J.}~\bibnamefont {Blumoff}}, \bibinfo
  {author} {\bibfnamefont {K.~M.}\ \bibnamefont {Sliwa}}, \bibinfo {author}
  {\bibfnamefont {L.}~\bibnamefont {Frunzio}}, \bibinfo {author} {\bibfnamefont
  {S.~M.}\ \bibnamefont {Girvin}}, \bibinfo {author} {\bibfnamefont
  {L.}~\bibnamefont {Jiang}}, \bibinfo {author} {\bibfnamefont
  {M.}~\bibnamefont {Mirrahimi}}, \bibinfo {author} {\bibfnamefont {M.~H.}\
  \bibnamefont {Devoret}},\ and\ \bibinfo {author} {\bibfnamefont {R.~J.}\
  \bibnamefont {Schoelkopf}},\ }\bibfield  {title} {\bibinfo {title} {A
  schr{\"o}dinger cat living in two boxes},\ }\href
  {https://doi.org/10.1126/science.aaf2941} {\bibfield  {journal} {\bibinfo
  {journal} {Science}\ }\textbf {\bibinfo {volume} {352}},\ \bibinfo {pages}
  {1087} (\bibinfo {year} {2016})}\BibitemShut {NoStop}%
\bibitem [{\citenamefont {Minganti}\ \emph {et~al.}(2016)\citenamefont
  {Minganti}, \citenamefont {Bartolo}, \citenamefont {Lolli}, \citenamefont
  {Casteels},\ and\ \citenamefont {Ciuti}}]{exacsolutionSciRep}%
  \BibitemOpen
  \bibfield  {author} {\bibinfo {author} {\bibfnamefont {F.}~\bibnamefont
  {Minganti}}, \bibinfo {author} {\bibfnamefont {N.}~\bibnamefont {Bartolo}},
  \bibinfo {author} {\bibfnamefont {J.}~\bibnamefont {Lolli}}, \bibinfo
  {author} {\bibfnamefont {W.}~\bibnamefont {Casteels}},\ and\ \bibinfo
  {author} {\bibfnamefont {C.}~\bibnamefont {Ciuti}},\ }\bibfield  {title}
  {\bibinfo {title} {Exact results for schr{\"o}dinger cats in
  driven-dissipative systems and their feedback control},\ }\href
  {https://doi.org/10.1038/srep26987} {\bibfield  {journal} {\bibinfo
  {journal} {Scientific Reports}\ }\textbf {\bibinfo {volume} {6}},\ \bibinfo
  {pages} {26987 EP } (\bibinfo {year} {2016})},\ \bibinfo {note}
  {article}\BibitemShut {NoStop}%
\bibitem [{\citenamefont {Goto}(2016{\natexlab{a}})}]{Goto16}%
  \BibitemOpen
  \bibfield  {author} {\bibinfo {author} {\bibfnamefont {H.}~\bibnamefont
  {Goto}},\ }\bibfield  {title} {\bibinfo {title} {Universal quantum
  computation with a nonlinear oscillator network},\ }\href
  {https://doi.org/10.1103/PhysRevA.93.050301} {\bibfield  {journal} {\bibinfo
  {journal} {Phys. Rev. A}\ }\textbf {\bibinfo {volume} {93}},\ \bibinfo
  {pages} {050301(R)} (\bibinfo {year} {2016}{\natexlab{a}})}\BibitemShut
  {NoStop}%
\bibitem [{\citenamefont {Goto}(2016{\natexlab{b}})}]{Goto2016}%
  \BibitemOpen
  \bibfield  {author} {\bibinfo {author} {\bibfnamefont {H.}~\bibnamefont
  {Goto}},\ }\bibfield  {title} {\bibinfo {title} {Bifurcation-based adiabatic
  quantum computation with a nonlinear oscillator network},\ }\href
  {https://doi.org/10.1038/srep21686} {\bibfield  {journal} {\bibinfo
  {journal} {Scientific Reports}\ }\textbf {\bibinfo {volume} {6}},\ \bibinfo
  {pages} {21686} (\bibinfo {year} {2016}{\natexlab{b}})}\BibitemShut {NoStop}%
\bibitem [{\citenamefont {Nigg}\ \emph {et~al.}(2017)\citenamefont {Nigg},
  \citenamefont {L{\"o}rch},\ and\ \citenamefont {Tiwari}}]{Nigg17}%
  \BibitemOpen
  \bibfield  {author} {\bibinfo {author} {\bibfnamefont {S.~E.}\ \bibnamefont
  {Nigg}}, \bibinfo {author} {\bibfnamefont {N.}~\bibnamefont {L{\"o}rch}},\
  and\ \bibinfo {author} {\bibfnamefont {R.~P.}\ \bibnamefont {Tiwari}},\
  }\bibfield  {title} {\bibinfo {title} {Robust quantum optimizer with full
  connectivity},\ }\href {https://doi.org/10.1126/sciadv.1602273} {\bibfield
  {journal} {\bibinfo  {journal} {Science Advances}\ }\textbf {\bibinfo
  {volume} {3}},\ \bibinfo {pages} {e1602273} (\bibinfo {year}
  {2017})}\BibitemShut {NoStop}%
\bibitem [{\citenamefont {Puri}\ \emph {et~al.}(2017)\citenamefont {Puri},
  \citenamefont {Andersen}, \citenamefont {Grimsmo},\ and\ \citenamefont
  {Blais}}]{Puri2017}%
  \BibitemOpen
  \bibfield  {author} {\bibinfo {author} {\bibfnamefont {S.}~\bibnamefont
  {Puri}}, \bibinfo {author} {\bibfnamefont {C.~K.}\ \bibnamefont {Andersen}},
  \bibinfo {author} {\bibfnamefont {A.~L.}\ \bibnamefont {Grimsmo}},\ and\
  \bibinfo {author} {\bibfnamefont {A.}~\bibnamefont {Blais}},\ }\bibfield
  {title} {\bibinfo {title} {Quantum annealing with all-to-all connected
  nonlinear oscillators},\ }\href {https://doi.org/10.1038/ncomms15785}
  {\bibfield  {journal} {\bibinfo  {journal} {Nature Communications}\ }\textbf
  {\bibinfo {volume} {8}},\ \bibinfo {pages} {15785} (\bibinfo {year}
  {2017})}\BibitemShut {NoStop}%
\bibitem [{\citenamefont {Verstraelen}\ and\ \citenamefont
  {Wouters}(2018)}]{gaussianmethod}%
  \BibitemOpen
  \bibfield  {author} {\bibinfo {author} {\bibfnamefont {W.}~\bibnamefont
  {Verstraelen}}\ and\ \bibinfo {author} {\bibfnamefont {M.}~\bibnamefont
  {Wouters}},\ }\bibfield  {title} {\bibinfo {title} {Gaussian quantum
  trajectories for the variational simulation of open quantum-optical
  systems},\ }\href {https://doi.org/10.3390/app8091427} {\bibfield  {journal}
  {\bibinfo  {journal} {Applied Sciences}\ }\textbf {\bibinfo {volume} {8}},\
  \bibinfo {pages} {1427} (\bibinfo {year} {2018})}\BibitemShut {NoStop}%
\bibitem [{\citenamefont {Buca}\ \emph {et~al.}(2019)\citenamefont {Buca},
  \citenamefont {Tindall},\ and\ \citenamefont {Jaksch}}]{Buca2019}%
  \BibitemOpen
  \bibfield  {author} {\bibinfo {author} {\bibfnamefont {B.}~\bibnamefont
  {Buca}}, \bibinfo {author} {\bibfnamefont {J.}~\bibnamefont {Tindall}},\ and\
  \bibinfo {author} {\bibfnamefont {D.}~\bibnamefont {Jaksch}},\ }\bibfield
  {title} {\bibinfo {title} {Non-stationary coherent quantum many-body dynamics
  through dissipation},\ }\href {https://doi.org/10.1038/s41467-019-09757-y}
  {\bibfield  {journal} {\bibinfo  {journal} {Nature Communications}\ }\textbf
  {\bibinfo {volume} {10}},\ \bibinfo {pages} {1730} (\bibinfo {year}
  {2019})}\BibitemShut {NoStop}%
\bibitem [{\citenamefont {Albert}\ \emph {et~al.}(2016)\citenamefont {Albert},
  \citenamefont {Bradlyn}, \citenamefont {Fraas},\ and\ \citenamefont
  {Jiang}}]{Albert2016}%
  \BibitemOpen
  \bibfield  {author} {\bibinfo {author} {\bibfnamefont {V.~V.}\ \bibnamefont
  {Albert}}, \bibinfo {author} {\bibfnamefont {B.}~\bibnamefont {Bradlyn}},
  \bibinfo {author} {\bibfnamefont {M.}~\bibnamefont {Fraas}},\ and\ \bibinfo
  {author} {\bibfnamefont {L.}~\bibnamefont {Jiang}},\ }\bibfield  {title}
  {\bibinfo {title} {Geometry and response of lindbladians},\ }\href
  {https://doi.org/10.1103/PhysRevX.6.041031} {\bibfield  {journal} {\bibinfo
  {journal} {Phys. Rev. X}\ }\textbf {\bibinfo {volume} {6}},\ \bibinfo {pages}
  {041031} (\bibinfo {year} {2016})}\BibitemShut {NoStop}%
\bibitem [{\citenamefont {Spohn}(1976)}]{Spohn1976}%
  \BibitemOpen
  \bibfield  {author} {\bibinfo {author} {\bibfnamefont {H.}~\bibnamefont
  {Spohn}},\ }\bibfield  {title} {\bibinfo {title} {Approach to equilibrium for
  completely positive dynamical semigroups of n-level systems},\ }\href
  {https://doi.org/https://doi.org/10.1016/0034-4877(76)90040-9} {\bibfield
  {journal} {\bibinfo  {journal} {Reports on Mathematical Physics}\ }\textbf
  {\bibinfo {volume} {10}},\ \bibinfo {pages} {189 } (\bibinfo {year}
  {1976})}\BibitemShut {NoStop}%
\bibitem [{\citenamefont {Spohn}(1977)}]{Spohn1977}%
  \BibitemOpen
  \bibfield  {author} {\bibinfo {author} {\bibfnamefont {H.}~\bibnamefont
  {Spohn}},\ }\bibfield  {title} {\bibinfo {title} {An algebraic condition for
  the approach to equilibrium of an open n-level system},\ }\href
  {https://doi.org/10.1007/BF00420668} {\bibfield  {journal} {\bibinfo
  {journal} {Letters in Mathematical Physics}\ }\textbf {\bibinfo {volume}
  {2}},\ \bibinfo {pages} {33} (\bibinfo {year} {1977})}\BibitemShut {NoStop}%
\bibitem [{\citenamefont {Nigro}(2019)}]{Nigro_2019}%
  \BibitemOpen
  \bibfield  {author} {\bibinfo {author} {\bibfnamefont {D.}~\bibnamefont
  {Nigro}},\ }\bibfield  {title} {\bibinfo {title} {On the uniqueness of the
  steady-state solution of the
  lindblad{\textendash}gorini{\textendash}kossakowski{\textendash}sudarshan
  equation},\ }\href {https://doi.org/10.1088/1742-5468/ab0c1c} {\bibfield
  {journal} {\bibinfo  {journal} {Journal of Statistical Mechanics: Theory and
  Experiment}\ }\textbf {\bibinfo {volume} {2019}},\ \bibinfo {pages} {043202}
  (\bibinfo {year} {2019})}\BibitemShut {NoStop}%
\bibitem [{\citenamefont {Rota}\ and\ \citenamefont
  {Savona}(2019)}]{antiferromagnet}%
  \BibitemOpen
  \bibfield  {author} {\bibinfo {author} {\bibfnamefont {R.}~\bibnamefont
  {Rota}}\ and\ \bibinfo {author} {\bibfnamefont {V.}~\bibnamefont {Savona}},\
  }\bibfield  {title} {\bibinfo {title} {Simulating frustrated antiferromagnets
  with quadratically driven qed cavities},\ }\href
  {https://doi.org/10.1103/PhysRevA.100.013838} {\bibfield  {journal} {\bibinfo
   {journal} {Phys. Rev. A}\ }\textbf {\bibinfo {volume} {100}},\ \bibinfo
  {pages} {013838} (\bibinfo {year} {2019})}\BibitemShut {NoStop}%
\bibitem [{\citenamefont {Dutta}\ \emph {et~al.}(2015)\citenamefont {Dutta},
  \citenamefont {Aeppli}, \citenamefont {Chakrabarti}, \citenamefont
  {Divakaran}, \citenamefont {Rosenbaum},\ and\ \citenamefont
  {Sen}}]{dutta2015}%
  \BibitemOpen
  \bibfield  {author} {\bibinfo {author} {\bibfnamefont {A.}~\bibnamefont
  {Dutta}}, \bibinfo {author} {\bibfnamefont {G.}~\bibnamefont {Aeppli}},
  \bibinfo {author} {\bibfnamefont {B.~K.}\ \bibnamefont {Chakrabarti}},
  \bibinfo {author} {\bibfnamefont {U.}~\bibnamefont {Divakaran}}, \bibinfo
  {author} {\bibfnamefont {T.~F.}\ \bibnamefont {Rosenbaum}},\ and\ \bibinfo
  {author} {\bibfnamefont {D.}~\bibnamefont {Sen}},\ }\href
  {https://doi.org/10.1017/CBO9781107706057} {\emph {\bibinfo {title} {Quantum
  Phase Transitions in Transverse Field Spin Models: From Statistical Physics
  to Quantum Information}}}\ (\bibinfo  {publisher} {Cambridge University
  Press},\ \bibinfo {year} {2015})\BibitemShut {NoStop}%
\bibitem [{\citenamefont {Finazzi}\ \emph {et~al.}(2015)\citenamefont
  {Finazzi}, \citenamefont {Le~Boit\'e}, \citenamefont {Storme}, \citenamefont
  {Baksic},\ and\ \citenamefont {Ciuti}}]{cornerspace}%
  \BibitemOpen
  \bibfield  {author} {\bibinfo {author} {\bibfnamefont {S.}~\bibnamefont
  {Finazzi}}, \bibinfo {author} {\bibfnamefont {A.}~\bibnamefont {Le~Boit\'e}},
  \bibinfo {author} {\bibfnamefont {F.}~\bibnamefont {Storme}}, \bibinfo
  {author} {\bibfnamefont {A.}~\bibnamefont {Baksic}},\ and\ \bibinfo {author}
  {\bibfnamefont {C.}~\bibnamefont {Ciuti}},\ }\bibfield  {title} {\bibinfo
  {title} {Corner-space renormalization method for driven-dissipative
  two-dimensional correlated systems},\ }\href
  {https://doi.org/10.1103/PhysRevLett.115.080604} {\bibfield  {journal}
  {\bibinfo  {journal} {Phys. Rev. Lett.}\ }\textbf {\bibinfo {volume} {115}},\
  \bibinfo {pages} {080604} (\bibinfo {year} {2015})}\BibitemShut {NoStop}%
\bibitem [{\citenamefont {Verstraelen}\ and\ \citenamefont
  {Wouters}(2019)}]{photoncondensate}%
  \BibitemOpen
  \bibfield  {author} {\bibinfo {author} {\bibfnamefont {W.}~\bibnamefont
  {Verstraelen}}\ and\ \bibinfo {author} {\bibfnamefont {M.}~\bibnamefont
  {Wouters}},\ }\bibfield  {title} {\bibinfo {title} {Temporal coherence of a
  photon condensate: A quantum trajectory description},\ }\href
  {https://doi.org/10.1103/PhysRevA.100.013804} {\bibfield  {journal} {\bibinfo
   {journal} {Phys. Rev. A}\ }\textbf {\bibinfo {volume} {100}},\ \bibinfo
  {pages} {013804} (\bibinfo {year} {2019})}\BibitemShut {NoStop}%
\bibitem [{\citenamefont {M{\o}lmer}\ \emph {et~al.}(1993)\citenamefont
  {M{\o}lmer}, \citenamefont {Castin},\ and\ \citenamefont
  {Dalibard}}]{MolmerJOSAB93}%
  \BibitemOpen
  \bibfield  {author} {\bibinfo {author} {\bibfnamefont {K.}~\bibnamefont
  {M{\o}lmer}}, \bibinfo {author} {\bibfnamefont {Y.}~\bibnamefont {Castin}},\
  and\ \bibinfo {author} {\bibfnamefont {J.}~\bibnamefont {Dalibard}},\
  }\bibfield  {title} {\bibinfo {title} {Monte carlo wave-function method in
  quantum optics},\ }\href {https://doi.org/10.1364/JOSAB.10.000524} {\bibfield
   {journal} {\bibinfo  {journal} {J. Opt. Soc. Am. B}\ }\textbf {\bibinfo
  {volume} {10}},\ \bibinfo {pages} {524} (\bibinfo {year} {1993})}\BibitemShut
  {NoStop}%
\bibitem [{\citenamefont {Haroche}\ and\ \citenamefont
  {Raimond}(2013)}]{haroche_raimond_2013}%
  \BibitemOpen
  \bibfield  {author} {\bibinfo {author} {\bibfnamefont {S.}~\bibnamefont
  {Haroche}}\ and\ \bibinfo {author} {\bibfnamefont {J.-M.}\ \bibnamefont
  {Raimond}},\ }\href@noop {} {\emph {\bibinfo {title} {Exploring the quantum:
  atoms, cavities, and photons}}}\ (\bibinfo  {publisher} {Oxford University
  Press},\ \bibinfo {year} {2013})\BibitemShut {NoStop}%
\bibitem [{\citenamefont {Carmichael}(2007)}]{CarmichaelBOOK}%
  \BibitemOpen
  \bibfield  {author} {\bibinfo {author} {\bibfnamefont {H.}~\bibnamefont
  {Carmichael}},\ }\href {https://books.google.fr/books?id=xgxOYkxW8JoC} {\emph
  {\bibinfo {title} {Statistical Methods in Quantum Optics 2: Non-Classical
  Fields}}},\ Theoretical and Mathematical Physics\ (\bibinfo  {publisher}
  {Springer Berlin Heidelberg},\ \bibinfo {year} {2007})\BibitemShut {NoStop}%
\bibitem [{\citenamefont {Wiseman}\ and\ \citenamefont
  {Milburn}(2010)}]{WisemanBOOK}%
  \BibitemOpen
  \bibfield  {author} {\bibinfo {author} {\bibfnamefont {H.}~\bibnamefont
  {Wiseman}}\ and\ \bibinfo {author} {\bibfnamefont {G.}~\bibnamefont
  {Milburn}},\ }\href {https://books.google.fr/books?id=ZNjvHaH8qA4C} {\emph
  {\bibinfo {title} {Quantum Measurement and Control}}}\ (\bibinfo  {publisher}
  {Cambridge University Press},\ \bibinfo {year} {2010})\BibitemShut {NoStop}%
\bibitem [{\citenamefont {Daley}(2014)}]{Daley14}%
  \BibitemOpen
  \bibfield  {author} {\bibinfo {author} {\bibfnamefont {A.~J.}\ \bibnamefont
  {Daley}},\ }\bibfield  {title} {\bibinfo {title} {Quantum trajectories and
  open many-body quantum systems},\ }\href
  {https://doi.org/10.1080/00018732.2014.933502} {\bibfield  {journal}
  {\bibinfo  {journal} {Advances in Physics}\ }\textbf {\bibinfo {volume}
  {63}},\ \bibinfo {pages} {77} (\bibinfo {year} {2014})},\ \Eprint
  {https://arxiv.org/abs/https://doi.org/10.1080/00018732.2014.933502}
  {https://doi.org/10.1080/00018732.2014.933502} \BibitemShut {NoStop}%
\bibitem [{\citenamefont {Weedbrook}\ \emph {et~al.}(2012)\citenamefont
  {Weedbrook}, \citenamefont {Pirandola}, \citenamefont {Garc\'{\i}a-Patr\'on},
  \citenamefont {Cerf}, \citenamefont {Ralph}, \citenamefont {Shapiro},\ and\
  \citenamefont {Lloyd}}]{GaussquantinfRMP}%
  \BibitemOpen
  \bibfield  {author} {\bibinfo {author} {\bibfnamefont {C.}~\bibnamefont
  {Weedbrook}}, \bibinfo {author} {\bibfnamefont {S.}~\bibnamefont
  {Pirandola}}, \bibinfo {author} {\bibfnamefont {R.}~\bibnamefont
  {Garc\'{\i}a-Patr\'on}}, \bibinfo {author} {\bibfnamefont {N.~J.}\
  \bibnamefont {Cerf}}, \bibinfo {author} {\bibfnamefont {T.~C.}\ \bibnamefont
  {Ralph}}, \bibinfo {author} {\bibfnamefont {J.~H.}\ \bibnamefont {Shapiro}},\
  and\ \bibinfo {author} {\bibfnamefont {S.}~\bibnamefont {Lloyd}},\ }\bibfield
   {title} {\bibinfo {title} {Gaussian quantum information},\ }\href
  {https://doi.org/10.1103/RevModPhys.84.621} {\bibfield  {journal} {\bibinfo
  {journal} {Rev. Mod. Phys.}\ }\textbf {\bibinfo {volume} {84}},\ \bibinfo
  {pages} {621} (\bibinfo {year} {2012})}\BibitemShut {NoStop}%
\bibitem [{\citenamefont {Ferraro}\ \emph {et~al.}(2005)\citenamefont
  {Ferraro}, \citenamefont {Olivares},\ and\ \citenamefont
  {Paris}}]{GaussQInf}%
  \BibitemOpen
  \bibfield  {author} {\bibinfo {author} {\bibfnamefont {A.}~\bibnamefont
  {Ferraro}}, \bibinfo {author} {\bibfnamefont {S.}~\bibnamefont {Olivares}},\
  and\ \bibinfo {author} {\bibfnamefont {M.}~\bibnamefont {Paris}},\
  }\href@noop {} {\emph {\bibinfo {title} {Gaussian States in Quantum
  Information}}},\ Napoli Series on physics and Astrophysics\ (\bibinfo
  {publisher} {Bibliopolis},\ \bibinfo {year} {2005})\ \bibinfo {note}
  {arXiv:quant-ph/0503237}\BibitemShut {NoStop}%
\bibitem [{\citenamefont {Bartolo}\ \emph {et~al.}(2017)\citenamefont
  {Bartolo}, \citenamefont {Minganti}, \citenamefont {Lolli},\ and\
  \citenamefont {Ciuti}}]{2photdrivingunraveling}%
  \BibitemOpen
  \bibfield  {author} {\bibinfo {author} {\bibfnamefont {N.}~\bibnamefont
  {Bartolo}}, \bibinfo {author} {\bibfnamefont {F.}~\bibnamefont {Minganti}},
  \bibinfo {author} {\bibfnamefont {J.}~\bibnamefont {Lolli}},\ and\ \bibinfo
  {author} {\bibfnamefont {C.}~\bibnamefont {Ciuti}},\ }\bibfield  {title}
  {\bibinfo {title} {Homodyne versus photon-counting quantum trajectories for
  dissipative kerr resonators with two-photon driving},\ }\href
  {https://doi.org/10.1140/epjst/e2016-60385-8} {\bibfield  {journal} {\bibinfo
   {journal} {The European Physical Journal Special Topics}\ }\textbf {\bibinfo
  {volume} {226}},\ \bibinfo {pages} {2705} (\bibinfo {year}
  {2017})}\BibitemShut {NoStop}%
\bibitem [{\citenamefont {Gardiner}\ and\ \citenamefont
  {Zoller}(2004)}]{quantumnoise}%
  \BibitemOpen
  \bibfield  {author} {\bibinfo {author} {\bibfnamefont {C.}~\bibnamefont
  {Gardiner}}\ and\ \bibinfo {author} {\bibfnamefont {P.}~\bibnamefont
  {Zoller}},\ }\href
  {https://www.amazon.com/Quantum-Noise-Non-Markovian-Applications-Synergetics/dp/3540223010?SubscriptionId=0JYN1NVW651KCA56C102&tag=techkie-20&linkCode=xm2&camp=2025&creative=165953&creativeASIN=3540223010}
  {\emph {\bibinfo {title} {Quantum Noise: A Handbook of Markovian and
  Non-Markovian Quantum Stochastic Methods with Applications to Quantum Optics
  (Springer Series in Synergetics)}}}\ (\bibinfo  {publisher} {Springer},\
  \bibinfo {year} {2004})\BibitemShut {NoStop}%
\bibitem [{\citenamefont {Breuer}\ and\ \citenamefont
  {Petruccioni}(2002)}]{breuer}%
  \BibitemOpen
  \bibfield  {author} {\bibinfo {author} {\bibfnamefont {H.-P.}\ \bibnamefont
  {Breuer}}\ and\ \bibinfo {author} {\bibfnamefont {F.}~\bibnamefont
  {Petruccioni}},\ }\href@noop {} {\emph {\bibinfo {title} {The theory of open
  Quantum Systems}}}\ (\bibinfo  {publisher} {Oxford university press},\
  \bibinfo {year} {2002})\BibitemShut {NoStop}%
\bibitem [{\citenamefont {Strunz}\ and\ \citenamefont
  {Percival}(1998)}]{Strunz_1998}%
  \BibitemOpen
  \bibfield  {author} {\bibinfo {author} {\bibfnamefont {W.~T.}\ \bibnamefont
  {Strunz}}\ and\ \bibinfo {author} {\bibfnamefont {I.~C.}\ \bibnamefont
  {Percival}},\ }\bibfield  {title} {\bibinfo {title} {Classical mechanics from
  quantum state diffusion - a phase-space approach},\ }\href
  {https://doi.org/10.1088/0305-4470/31/7/014} {\bibfield  {journal} {\bibinfo
  {journal} {Journal of Physics A: Mathematical and General}\ }\textbf
  {\bibinfo {volume} {31}},\ \bibinfo {pages} {1801} (\bibinfo {year}
  {1998})}\BibitemShut {NoStop}%
\bibitem [{\citenamefont {Link}\ \emph {et~al.}(2019)\citenamefont {Link},
  \citenamefont {Luoma},\ and\ \citenamefont
  {Strunz}}]{phasetransitionwithtrajectories}%
  \BibitemOpen
  \bibfield  {author} {\bibinfo {author} {\bibfnamefont {V.}~\bibnamefont
  {Link}}, \bibinfo {author} {\bibfnamefont {K.}~\bibnamefont {Luoma}},\ and\
  \bibinfo {author} {\bibfnamefont {W.~T.}\ \bibnamefont {Strunz}},\ }\bibfield
   {title} {\bibinfo {title} {Revealing the nature of nonequilibrium phase
  transitions with quantum trajectories},\ }\href
  {https://doi.org/10.1103/PhysRevA.99.062120} {\bibfield  {journal} {\bibinfo
  {journal} {Phys. Rev. A}\ }\textbf {\bibinfo {volume} {99}},\ \bibinfo
  {pages} {062120} (\bibinfo {year} {2019})}\BibitemShut {NoStop}%
\bibitem [{sup()}]{sup}%
  \BibitemOpen
  \href@noop {} {}\bibinfo {note} {See Supplemental Material at [URL will be
  inserted by publisher] for the full GTA equations, a benchmark of the
  numerical performance, a reproduction of the phase transition in the quantum
  regime and benchmark studies on a dimer}\BibitemShut {NoStop}%
\bibitem [{\citenamefont {del Campo}\ and\ \citenamefont
  {Zurek}(2014)}]{KZreview}%
  \BibitemOpen
  \bibfield  {author} {\bibinfo {author} {\bibfnamefont {A.}~\bibnamefont {del
  Campo}}\ and\ \bibinfo {author} {\bibfnamefont {W.~H.}\ \bibnamefont
  {Zurek}},\ }\bibfield  {title} {\bibinfo {title} {Universality of phase
  transition dynamics: Topological defects from symmetry breaking},\ }\href
  {https://doi.org/10.1142/S0217751X1430018X} {\bibfield  {journal} {\bibinfo
  {journal} {International Journal of Modern Physics A}\ }\textbf {\bibinfo
  {volume} {29}},\ \bibinfo {pages} {1430018} (\bibinfo {year} {2014})},\
  \Eprint {https://arxiv.org/abs/https://doi.org/10.1142/S0217751X1430018X}
  {https://doi.org/10.1142/S0217751X1430018X} \BibitemShut {NoStop}%
\bibitem [{\citenamefont {Binder}(1981{\natexlab{a}})}]{Binder1981}%
  \BibitemOpen
  \bibfield  {author} {\bibinfo {author} {\bibfnamefont {K.}~\bibnamefont
  {Binder}},\ }\bibfield  {title} {\bibinfo {title} {Finite size scaling
  analysis of ising model block distribution functions},\ }\href
  {https://doi.org/10.1007/BF01293604} {\bibfield  {journal} {\bibinfo
  {journal} {Zeitschrift f{\"u}r Physik B Condensed Matter}\ }\textbf {\bibinfo
  {volume} {43}},\ \bibinfo {pages} {119} (\bibinfo {year}
  {1981}{\natexlab{a}})}\BibitemShut {NoStop}%
\bibitem [{\citenamefont {Binder}(1981{\natexlab{b}})}]{binderpar}%
  \BibitemOpen
  \bibfield  {author} {\bibinfo {author} {\bibfnamefont {K.}~\bibnamefont
  {Binder}},\ }\bibfield  {title} {\bibinfo {title} {Critical properties from
  monte carlo coarse graining and renormalization},\ }\href
  {https://doi.org/10.1103/PhysRevLett.47.693} {\bibfield  {journal} {\bibinfo
  {journal} {Phys. Rev. Lett.}\ }\textbf {\bibinfo {volume} {47}},\ \bibinfo
  {pages} {693} (\bibinfo {year} {1981}{\natexlab{b}})}\BibitemShut {NoStop}%
\bibitem [{\citenamefont {Biondi}\ \emph {et~al.}(2018)\citenamefont {Biondi},
  \citenamefont {Blatter},\ and\ \citenamefont {Schmidt}}]{Biondi18}%
  \BibitemOpen
  \bibfield  {author} {\bibinfo {author} {\bibfnamefont {M.}~\bibnamefont
  {Biondi}}, \bibinfo {author} {\bibfnamefont {G.}~\bibnamefont {Blatter}},\
  and\ \bibinfo {author} {\bibfnamefont {S.}~\bibnamefont {Schmidt}},\
  }\bibfield  {title} {\bibinfo {title} {Emergent light crystal from
  frustration and pump engineering},\ }\href
  {https://doi.org/10.1103/PhysRevB.98.104204} {\bibfield  {journal} {\bibinfo
  {journal} {Phys. Rev. B}\ }\textbf {\bibinfo {volume} {98}},\ \bibinfo
  {pages} {104204} (\bibinfo {year} {2018})}\BibitemShut {NoStop}%
\bibitem [{\citenamefont {Vicentini}\ \emph {et~al.}(2019)\citenamefont
  {Vicentini}, \citenamefont {Minganti}, \citenamefont {Biella}, \citenamefont
  {Orso},\ and\ \citenamefont {Ciuti}}]{Vicentini19}%
  \BibitemOpen
  \bibfield  {author} {\bibinfo {author} {\bibfnamefont {F.}~\bibnamefont
  {Vicentini}}, \bibinfo {author} {\bibfnamefont {F.}~\bibnamefont {Minganti}},
  \bibinfo {author} {\bibfnamefont {A.}~\bibnamefont {Biella}}, \bibinfo
  {author} {\bibfnamefont {G.}~\bibnamefont {Orso}},\ and\ \bibinfo {author}
  {\bibfnamefont {C.}~\bibnamefont {Ciuti}},\ }\bibfield  {title} {\bibinfo
  {title} {Optimal stochastic unraveling of disordered open quantum systems:
  Application to driven-dissipative photonic lattices},\ }\href
  {https://doi.org/10.1103/PhysRevA.99.032115} {\bibfield  {journal} {\bibinfo
  {journal} {Phys. Rev. A}\ }\textbf {\bibinfo {volume} {99}},\ \bibinfo
  {pages} {032115} (\bibinfo {year} {2019})}\BibitemShut {NoStop}%
\bibitem [{\citenamefont {Verstraelen}\ and\ \citenamefont
  {Wouters}(2020)}]{dynamicalPT}%
  \BibitemOpen
  \bibfield  {author} {\bibinfo {author} {\bibfnamefont {W.}~\bibnamefont
  {Verstraelen}}\ and\ \bibinfo {author} {\bibfnamefont {M.}~\bibnamefont
  {Wouters}},\ }\bibfield  {title} {\bibinfo {title} {Classical critical
  dynamics in quadratically driven kerr resonators},\ }\href
  {https://doi.org/10.1103/PhysRevA.101.043826} {\bibfield  {journal} {\bibinfo
   {journal} {Phys. Rev. A}\ }\textbf {\bibinfo {volume} {101}},\ \bibinfo
  {pages} {043826} (\bibinfo {year} {2020})}\BibitemShut {NoStop}%
\end{thebibliography}

\providecommand{\noopsort}[1]{}\providecommand{\singleletter}[1]{#1}%

\end{document}


\title{Supplementary material to \emph{Gaussian trajectory approach to dissipative phase transitions: the case of quadratically driven photonic lattices}}

\author{Wouter Verstraelen}
\email{Wouter.Verstraelen@uantwerpen.be}
\affiliation{TQC, University of Antwerp, B-2610 Wilrijk, Belgium}

\author{Riccardo Rota}%
 \email{Riccardo.Rota@epfl.ch}
 \author{Vincenzo Savona}
\affiliation{Institute of Physics, Ecole Polytechnique F\'ed\'erale de Lausanne (EPFL), CH-1015 Lausanne, Switzerland}%

\author{Michiel Wouters}
\affiliation{TQC, University of Antwerp, B-2610 Wilrijk, Belgium}

\date{December 2019}

\maketitle
\onecolumngrid
\newpage
\section{Gaussian Trajectory equations}\label{sec:gausseq}

Following the approach of \cite{gaussianmethod}, the evolution of a generic expectation value $\ev*{\Oop}$ under heterodyne unraveling for all decay channels is given by
\begin{align}\label{eq:generictrajectorycorrelation}
    \dd\ev*{\Oop}=&i\ev{\comm{\hat{H}}{\Oop}}\dd{t}-\frac{1}{2}\sum_{j,k}\left(\ev{\acomm{\hat{\Gamma}_{j,k}^\dagger \hat{\Gamma}_{j,k}}{\Oop}}-2\ev{\hat{\Gamma}_{j,k}^\dagger\Oop\hat{\Gamma}_{j,k}}\right)\dd{t}\nonumber\\&+\sum_{j,k}\left(\ev{\hat{\Gamma}_{j,k}^\dagger(\Oop-\ev*{\Oop})}\dd{Z}_{j,k}+\ev{(\Oop-\ev*{\Oop})\hat{\Gamma}_{j,k}}\dd{Z}_{j,k}^*\right),
\end{align}
where $dZ_i=\frac{1}{\sqrt{2}}(dW_{x,i}+idW_{p,i})$ is a complex Wiener process satisfying $\abs{dZ_i}^2=dt$. By assuming a Gaussian ansatz, Wick decompositions are performed and and the trajectory is expressed entirely in terms of the first and second central moments $\alpha_n=\ev*{\aop_n}$, $u_{nm}:=\ev*{\delop_n\delop_m}=\ev*{\aop_n\aop_m}-\a{n}\a{m}$, $v_{nm}:=\ev*{\delop_n^\dagger\delop_m}=\ev*{\cop_n\aop_m}-\a{n}^*\a{m}$. Note that $\u{nm}=\u{mn}$ and $\v{nm}=\v{mn}^*$. Using \eqref{eq:generictrajectorycorrelation}, the evolution of Gaussian moments is given by

\begin{align}\label{eq:gausseq}
\dd \a{n}&=\left[\left(-\frac{\gamma}{2}+i\Delta\right)\a{n}+i\frac{J}{z}\sum_{n'}\a{n'}
-(\eta+Ui)(\abs{\a{n}}^2\a{n}+2\a{n}\v{nn}+\a{n}^*\u{nn})-iG\a{n}^*\right]\dd{t}\nonumber\\
&+\sqrt{\gamma}\sum_i\left(\v{in}\dZ{1}{i}+\u{in}{\dZ{1}{i}}^*\right)\nonumber\\
&+2\sqrt{\eta}\sum_i\left(\a{i}^*\v{in}\dZ{2}{i}+\a{i}\u{in}{\dZ{2}{i}}^*\right)\\
\dd \u{nm}&=\left[(-\gamma+2i\Delta)\u{nm}+i\frac{J}{z}\left(\sum_{n'}\u{n'm}+\sum_{m'}\u{nm'}\right)-iG(\v{nm}+\v{mn}+\delta_{n,m})\right.\nonumber\\
&\left.-(\eta+Ui)\left(\v{nm}(\a{n}^2+\u{nn})+\v{mn}(\a{m}^2+\u{mm})+2\u{nm}(\abs{\a{n}}^2+\abs{\a{m}}^2+\v{nn}+\v{mm})+\delta_{n,m}(\a{n}\a{m}+\u{nm})\right)\nonumber\right.\\
&\left.-\gamma\sum_i\left(\u{mi}\v{in}+\u{ni}\v{im}\right)\right.\nonumber\\
&\left.-4\eta\sum_i\abs{\a{i}}^2\left(\u{mi}\v{in}+\u{ni}\v{im}\right)\right]\dd{t}\nonumber\\
&+2\sqrt{\eta}\sum_i\left(\v{in}\v{im}\dZ{2}{i}+\u{in}\u{im}{\dZ{2}{i}}^*\right)\nonumber\\
\dd \v{nm}&=\left[iU\left(2\v{nm}(\abs{\a{n}}^2-\abs{\a{m}}^2+\v{nn}-\v{mm})+\u{nm}(\a{n}^{*2}+\u{nn}^*)-\u{nm}^*(\a{m}^2+\u{mm})\right)\right.\nonumber\\
&\left.-\eta\left(2\v{nm}(\abs{\a{n}}^2+\abs{\a{m}}^2+\v{nn}+\v{mm})+\u{nm}(\a{n}^{*2}+\u{nn}^*)+\u{nm}^*(\a{m}^2+\u{mm})\right)\right.\nonumber\\
&\left. -i\frac{J}{z}\left(\sum_{n'}\v{n'm}-\sum_{m'}\v{nm'}\right)
+iG(\u{nm}-\u{nm}^*)-\gamma\v{nm} \right.\nonumber\\
&\left.-\gamma\sum_i\left(\v{ni}\v{im}+\u{ni}^*\u{im}\right)\right.\nonumber\\
&\left.-4\eta\sum_i\abs{\a{i}}^2\left(\v{ni}\v{im}+\u{ni}^*\u{im}\right)\right]\dd{t}\nonumber\\
&+2\sqrt{\eta}\sum_i\left(\u{ni}^*\v{mi}^*\dZ{2}{i}+\v{ni}\u{im}{\dZ{2}{i}}^*\right)\nonumber,
\end{align}
where primed indices refer to nearest-neighbours only.

In Fig. \ref{fig:benchmarking} (a), we benchmark the numerical performance of the method as function of system size. The scaling of required resources goes quadratically, as was predicted in ref \cite{gaussianmethod}. We also observe that disregarding the two-photon loss ($\eta=0$) results in a speedup of order two. Indeed, about half of the terms in \eqref{eq:gausseq} are proportional to $\eta$. 

\section{Quantum phase transition at small loss rate}\label{sec:quantumPT}

In this section, we apply the GTA to the study the lattice of quadratically driven Kerr resonators in regimes where the one-photon losses is small respect to the Hamiltonian and we benchmark it with the results obtained in Ref. \cite{PRLspinmodel}. The purpose of this study is two-fold. At first, we want to demonstrate that the GTA is not limited to the description of classically correlated states in open quantum systems, but can be applied also in the study of dissipative phase transitions with a quantum criticality: indeed, the Gaussian ansatz may exhibit multimode entanglement, as shown also in several studies of continuous-variable quantum information \cite{GaussquantinfRMP}. Secondly, we want to show that two-photon losses do not play a relevant role in the emergence of the dissipative phase transition, considering the critical behavior of the system for $\eta = 0$.

\begin{figure}
\subfloat[]{\includegraphics[width=0.45\linewidth]{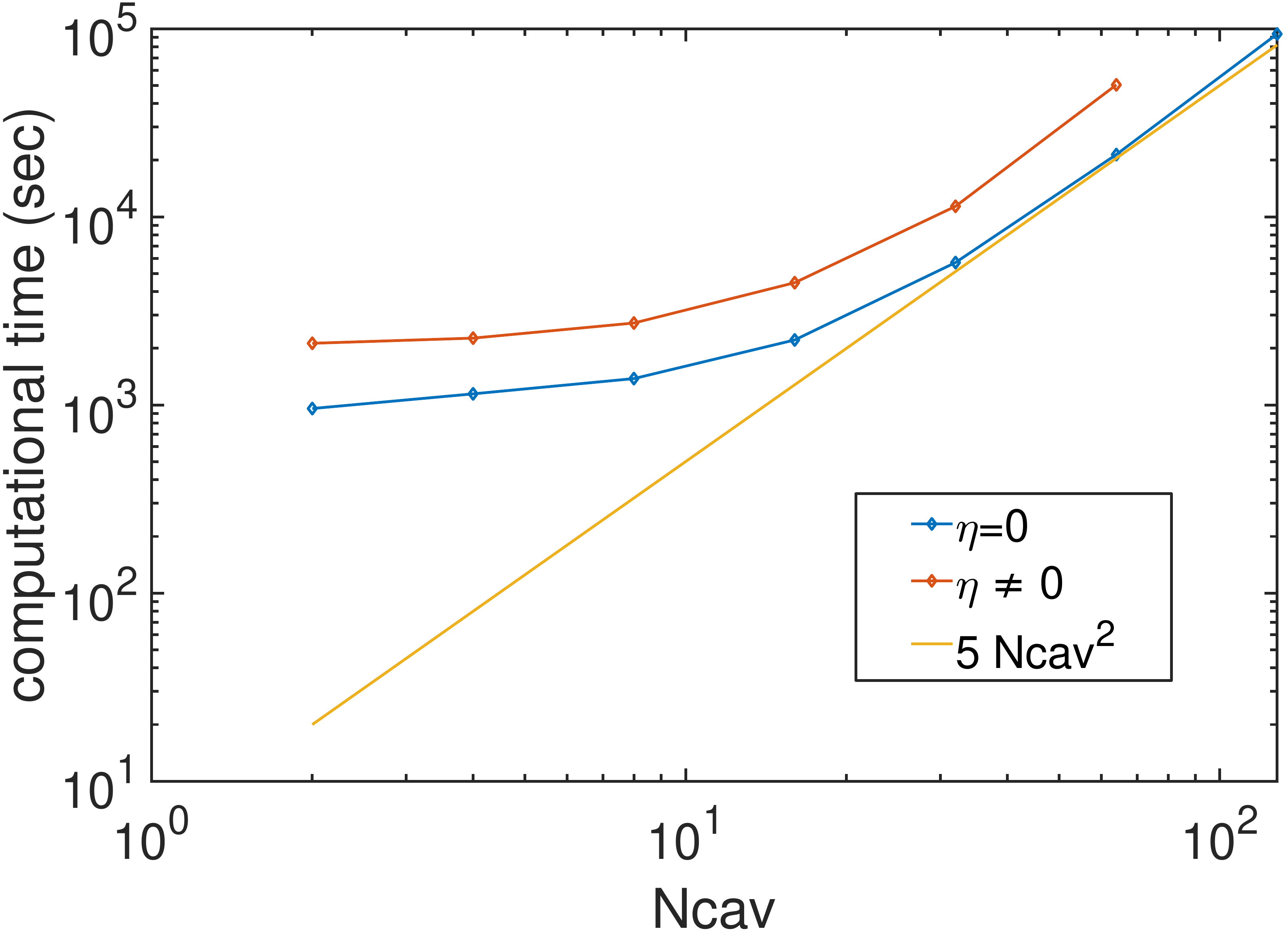}}\quad
\subfloat[]{\includegraphics[width=0.45\linewidth]{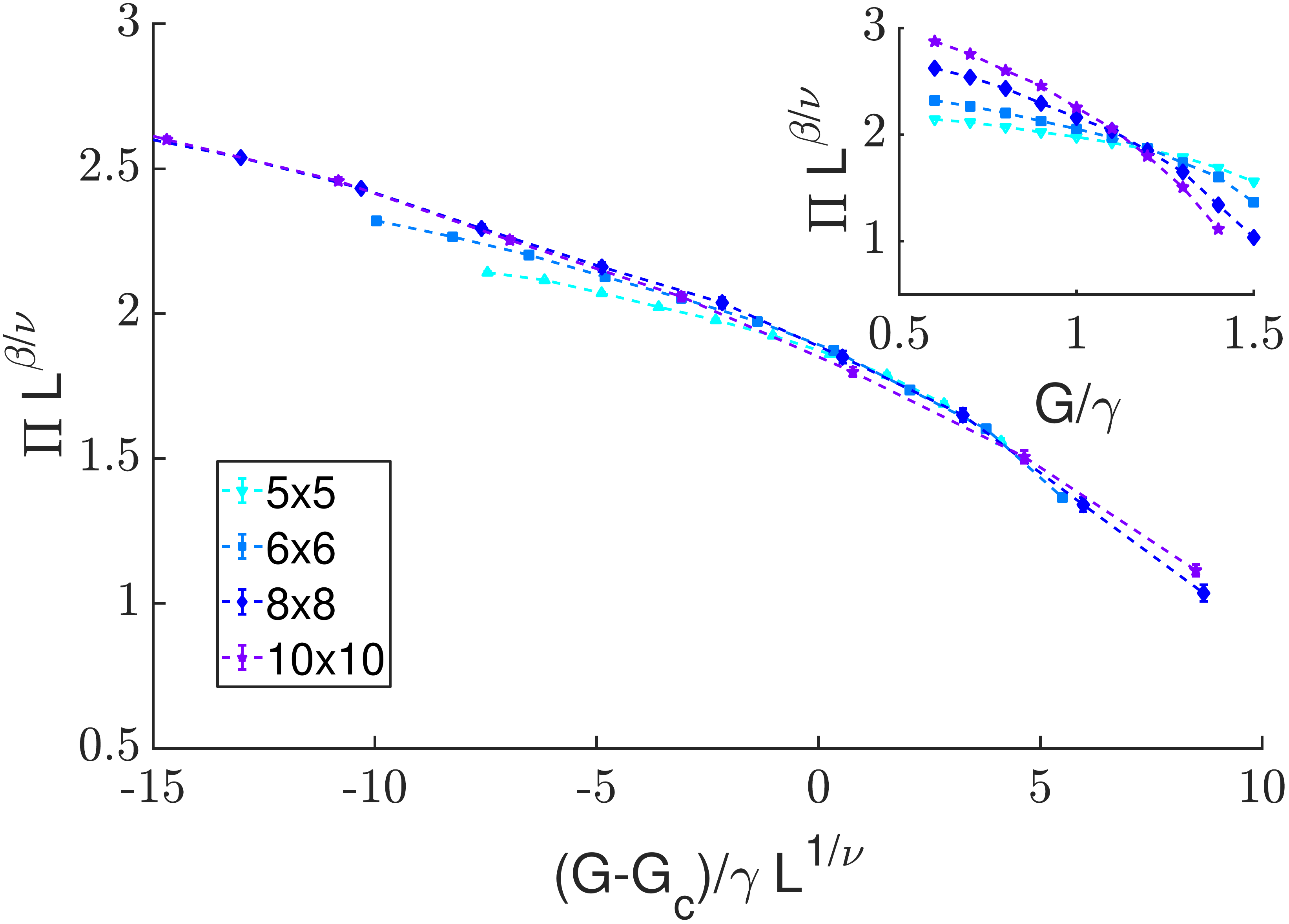}}
    \caption{(a) Computational time needed to recover 100 Gaussian trajectories on a desktop computer as function of system size. For every system size, the trajectories were evolved for 100 $\gamma^{-1}$ with numerical (Euler-Mayurama) timestep $h=10^{-4}\gamma^{-1}$. The quadratic scaling is evident (b) Parity obtained with gaussian trajectories for lattices of different sizes $(U=40\gamma,J=20\gamma,\Delta=-20\gamma,\eta=0)$. The critical exponents which allow to recover the universal behavior of the rescaled quanties are $\nu = 0.62997$ and $\beta = 0.32642$, i.e. the critical exponents for the correlation length and the order parameter of the 2D quantum transverse Ising model. Note that for Gaussian states, $\Pi$ is obtained by the expectation of the Wigner-function value in the origin of multimode phase-space  \cite{GaussQInf} .}
\label{fig:benchmarking}
\end{figure}

In Fig. \ref{fig:benchmarking}-(b), we show the results for the steady-state expectation value of the parity $\Pi=\ev{\exp{i\pi\cop\aop}}$, as obtained with the GTA, for the following set of parameters: $U=40\gamma, J=20\gamma, \Delta=-20\gamma, \eta=0$ (i.e. the same regime of parameter considered in Ref. \cite{PRLspinmodel}, except for the value of the two-photon loss rate). The results show the emergence of a critical point at $G_c \simeq 1.2\gamma$, with the critical exponents of the quantum transverse Ising model.

Note that the parity cannot be used to study the phase transition with classical criticality in the regimes of large loss rates, that we considered in the main text. In the latter case, indeed, the strong classical fluctuations mix the even and odd eigenstates of the steady-state density matrix at all values of $G$, such that $\Pi = 0$ on both sides of the phase transitions.

\section{Benchmark studies on a dimer}

\begin{figure}
    \centering
    \includegraphics[width=0.9\linewidth]{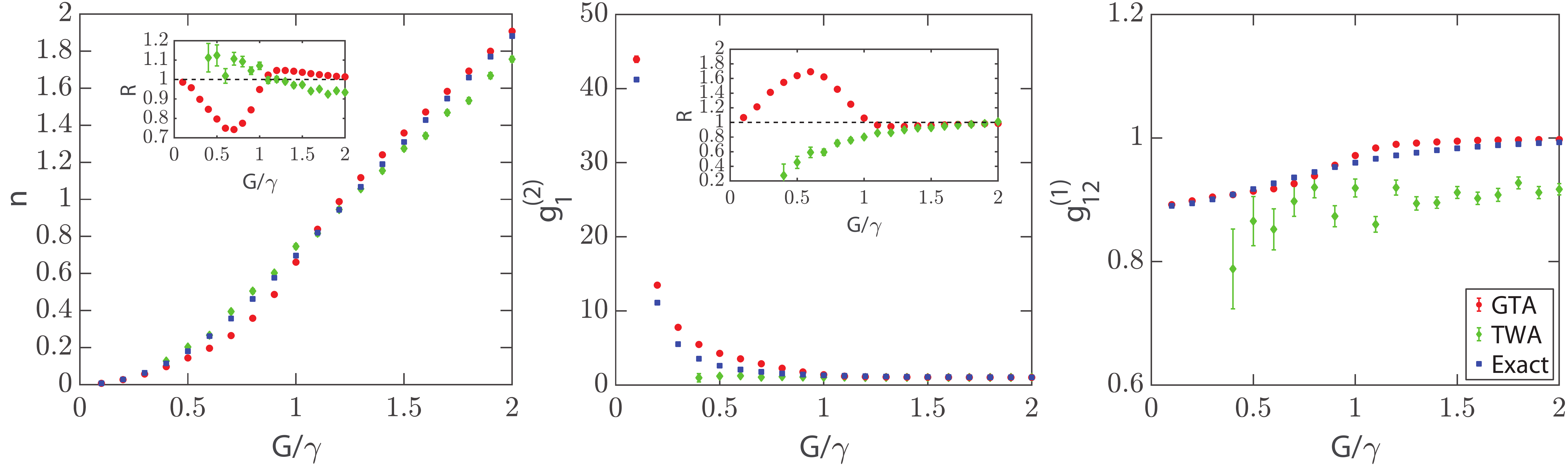}
    \caption{Predictions of the GTA-method in a dimer for $U=J=-\Delta=\gamma$.  The left two panels display average photon number and second-order correlation function in site 1. The results are compared with those obtained from the exact solution of the master equation of the system and also with the prediction of the truncated Wigner approximation (TWA). Both GTA and TWA provide a reasonable prediction for this parameter regime. On the right panel, we see however from the plot of  that the GTA method is, unlike the TWA, able to capture the correlations between modes properly, which is important to the study of critical phenomena.}
    \label{fig:dimerbenchmark}
\end{figure}

In this section, we consider the case of a quadratically driven-dissipative dimer - i.e. a system with $N=2$ sites, described by the same master equation as the extended lattice (Eq. 3 of the main text). In Fig. \ref{fig:dimerbenchmark}, we compare the predictions of the GTA method with the exact result obtained by solving the master equation in a truncated Fock basis. For the case of local observables, such as the photon number $n_1=\ev{\cop_1\aop_1}$ (left panel) and the local second-order correlation function $g^{(2)}_1=\ev{\cop_1\cop_1\aop_1\aop_1}/n_1^2$ (middle panel), we see that GTA is able to give accurate results in the limits of small and large values of $G/\gamma$, and we notice small deviations in the intermediate region of $G/\gamma \sim 1$, as already pointed out in Ref. \cite{gaussianmethod}. For the case of the non-local first-order correlation function $g^{(1)}_{12}=\ev{\cop_1\aop_2}/n_1$ (right panel), instead, we notice a good agreement between GTA and exact in the whole range of $G/\gamma$. This last result suggests that the GTA method is able to provide a reliable description of the non-local correlations which give rise to the emergence of the critical behavior in extended lattices.

In Fig. \ref{fig:dimerbenchmark}, we show also the results obtained for the steady-state expectation values obtained with a truncated Wigner approximation (TWA), a theoretical approach commonly used in the study of driven-dissipative systems \cite{qfloflight,*TWACarusottoCiuti,*TWAWoutersSavona} \eqref{eq:TWA}.  While we see that for the local observables, the accuracy of the predictions from both methods is comparable, the TWA method clearly fails in describing the correlations between sites, as opposed to the GTA.

\subsection{For comparison: truncated Wigner approximation}
Following the standard recipe of using operator correspondences \cite{quantumnoise} to translate the Master equation in a Fokker-Planck-like equation, neglecting derivatives of order higher than two and exploiting Feynman-Kac, the evolution equation
\begin{align}\label{eq:TWA}
  \dd{\aw^{(n)}}&=\left[\left(-\frac{\gamma}{2}+i\Delta\right)\aw^{(n)}+i\frac{J}{z}\sum_{n'}\aw^{(n')}-(\eta+Ui)(\abs*{\aw^{(n)}}^2-1)\aw^{(n)}-iG\aw^{(n)*}\right]\dd{t}+\sqrt{\frac{\gamma}{2}+2\eta\abs{\aw}^2}\dd{Z}_n,
\end{align}
is obtained for the symmetrized fields $\aw^{(n)}$ (note the similarity with \eqref{eq:gausseq}).


\providecommand{\noopsort}[1]{}\providecommand{\singleletter}[1]{#1}%